\documentclass[12pt,preprint]{aastex}

\newcommand{\kms}{$\rm km~s^{-1}$}
\newcommand{\masy}{$\rm mas~yr^{-1}$}
\newcommand{\logg}{$\rm \log g\ $}
\newcommand{\Teff}{$\rm T_{eff}\ $}
\newcommand{\loggns}{$\rm \log g$}
\newcommand{\Teffns}{$\rm T_{eff}$}
\newcommand{\Msun}{$\rm M_{\sun}$}

\slugcomment{To appear ApJ, 20May2004}

\shorttitle{SDSS White Dwarf Stars}
\shortauthors{Kleinman et al.}

\begin{document}

\title{A Catalog of Spectroscopically Identified White Dwarf Stars in the 
       First Data Release of the Sloan Digital Sky Survey}

\author{S. J. Kleinman\altaffilmark{1},
Hugh C. Harris\altaffilmark{2},
Daniel J. Eisenstein\altaffilmark{3},
James Liebert\altaffilmark{3},
Atsuko Nitta\altaffilmark{1},
Jurek Krzesi\'{n}ski\altaffilmark{1,4},
Jeffrey A. Munn\altaffilmark{2},
Conard C. Dahn\altaffilmark{2},
Suzanne L. Hawley\altaffilmark{5},
Jeffrey R. Pier\altaffilmark{2},
Gary Schmidt\altaffilmark{3},
Nicole M. Silvestri\altaffilmark{5},
J. Allyn Smith\altaffilmark{6,7},
Paula Szkody\altaffilmark{5},
Michael A. Strauss\altaffilmark{8},
G. R. Knapp\altaffilmark{8},
Matthew J. Collinge\altaffilmark{8},
A. S. Mukadam\altaffilmark{9},
D. Koester\altaffilmark{10},
Alan Uomoto\altaffilmark{11,12},
D. J. Schlegel\altaffilmark{8},
Scott F. Anderson\altaffilmark{5},
J. Brinkmann\altaffilmark{1},
D.Q. Lamb\altaffilmark{13},
Donald P. Schneider\altaffilmark{14},
and
Donald G. York\altaffilmark{13}
}
\altaffiltext{1}{New Mexico St. Univ., Apache Pt. Observatory, PO Box 59
                 Sunspot, NM 88349: sjnk@apo.nmsu.edu.}
\altaffiltext{2}{U.S. Naval Observatory, PO Box 1149, Flagstaff, AZ 86002.}
\altaffiltext{3}{Steward Observatory, Univ. of Arizona, 933 N. Cherry
                 Ave., Tucson, AZ 85721.}
\altaffiltext{4}{Mt. Suhora Observatory, Cracow Pedgagogical Univ., ul.
                 Podchor\c{a}\.{z}ych 2, 30--084 Cracow, Poland.}
\altaffiltext{5}{Dept. of Astronomy, Univ. of Washington, Box 351580, Seattle,
                 WA 98195.}
\altaffiltext{6}{Los Alamos Natl. Lab., NIS-4 MS-D448, Los Alamos, NM 87545.}
\altaffiltext{7}{Dept. Of Physics and Astronomy, Univ. of Wyoming, PO Box 3905,
                 Laramie, WY 82071.}
\altaffiltext{8}{Princeton Univ. Observatory, Princeton, NJ, 08544.}
\altaffiltext{9}{Dept. of Astronomy, Univ. of Texas, Austin, TX 78712.}
\altaffiltext{10}{Institut f\"ur Theoretische Physik und Astrophysik
                  Universit\"at Kiel, 24098  Kiel, Germany.}
\altaffiltext{11}{Dept. of Physics and Astronomy, John Hopkins Univ., 3400 N.
                  Charles St., Baltimore, MD, 21218.}
\altaffiltext{12}{Carnegie Observatories, 813 Santa Barbara St., Pasadena, 
                  CA, 91101.}
\altaffiltext{13}{Dept. of Astronomy and Astrophysics, Univ. of Chicago, 
                  5640 S. Ellis Ave., Chicago, IL, 60637.}
\altaffiltext{14}{Dept. of Astronomy and Astrophysics, Penn. St. Univ., 
                  University Park, PA, 16802.}

\begin{abstract}
We present the full spectroscopic white dwarf and hot subdwarf sample from
the SDSS first data release, DR1.   We find 2551 white dwarf stars of
various types, 240 hot subdwarf stars, and an additional 144 objects we
have identified as uncertain white dwarf stars.
Of the white dwarf stars, 1888 are non-magnetic
DA types and 171, non-magnetic DBs.
The remaining (492) objects consist of all different types of white dwarf stars:
DO, DQ, DC, DH, DZ, hybrid stars like DAB, etc., and those with non-degenerate
companions.
We fit the DA and DB spectra
with a grid of models to determine the \Teff and \logg for each object.
For all objects, we provide coordinates, proper motions, SDSS photometric
magnitudes, 
and enough information to retrieve the spectrum/image from the SDSS public
database.  This catalog nearly doubles the known sample of
spectroscopically-identified white dwarf stars.  In the DR1 imaged area of the
sky, we increase the known sample of white dwarf stars by a factor of 8.5.
We also comment on several particularly interesting objects in this sample.
\end{abstract}


\section{Introduction}

The Sloan Digital Sky Survey \citep[SDSS:][]{yor00} is a continuing
imaging and spectroscopic survey of some seven to ten thousand square degrees
in the north Galactic cap.  Although its main focus is extra-galactic,
there are many Galactic spin-off projects resulting from the survey.
The SDSS's principal science objectives focus on obtaining redshifts of distant
galaxies and quasars by first imaging the sky in 5 passbands, then selecting potential
targets for spectroscopic follow-up based on the 5-band photometry. This
spectroscopic selection process is referred to as ``targeting'' and many
different targeting categories, each with different criteria and
priorities, are used to fill all the fibers available on each 640-fiber
spectroscopic plate.  Where there are not enough primary targets
(ie., galaxies, QSOs, etc.) to fill  a given plate,
the lower priority targeting categories (which include various stellar and
serendipity categories) get to allocate fibers.
SDSS obtained spectra
are of high enough quality to allow accurate object and line identifications
well beyond those necessary for redshift determinations.  
We thus end up with high-quality stellar spectra from objects directly
targeted as interesting Galactic objects as well as those thought to be,
but which ultimately were not, extra-Galactic objects.

Complementary to \citet{har03} which presented white dwarf stars from  a sample
of early SDSS data, this paper reports on the white dwarf
stars found in the spectroscopic data contained within the SDSS Data Release 1 
\citep[ DR1: see also \url{http://www.sdss.org/dr1}]{dr1}.   In a 190 deg$^2$
area of sky, \citet{har03} found 260 white dwarf stars. In an area of sky 7.15
times larger, we find a factor of 9.85 more white dwarf stars, or an increased
density of approximately 38\% compared to that of \citet{har03}. We suspect
this difference is simply due to random fluctuations in how we target and
acquire white dwarf spectra in the SDSS. QSO target selection code changes,
for example, can have significant effects on the ultimate yield of SDSS white
dwarf spectra.

Since the DR1 spectroscopic coverage on the sky contains the area analyzed
in \citet{har03}, all those stars are included here, although perhaps with a
different spectrum than was analyzed in that work.
The white dwarf
sample presented here is not meant to be a statistically complete or even
well-defined sample of white dwarf stars; it merely represents the white dwarf
stars that happen to have spectra in DR1.  There are undoubtedly many more
white dwarf stars contained within DR1 for which we only have photometry.  We
make no attempt to report on those stars here (except see the Appendix for data
on previously known white dwarf stars in DR1 for which we do not have spectra).

We also include spectra
for hot subdwarf stars, the sdB and sdO stars, since they overlap the hot
white dwarf stars in color-color space and need to be identified in order
to find the white dwarf stars (besides being interesting in their own
right, of course).  For the  broader context of white dwarf stars in the SDSS
and particular notes of some unusual objects, see the \citet{har03} paper
and the earlier simulations of \citet{fan99} who discusses where the white dwarf
stars should appear in the SDSS photometric color space.

The SDSS spectra are well-described in \citet{edr} and \citet{dr1}.  Briefly,
they cover a wavelength range of approximately 3800--9200\AA\ with a resolution
$\sim 1800$ and are spectrophotometrically calibrated to within about 10\% on
average. The average S/N of a {\it g}=20.2 spectrum is $\sim 4$ per pixel
and the redshift accuracy is of order 30 \kms, as determined for the SDSS
galaxy sample.

The \citet{mcc99} catalog lists 2249 white dwarf stars while the online updates
(at {\url http://www.astronomy.villanova.edu/WDCatalog/index.htm}) now include a
total of 3066 white dwarf stars (as of June, 2003), some of which are included
in the first SDSS white dwarf catalog paper by \citet{har03}.  Here, we present
2551 certain white dwarf stars, 240 hot subdwarf stars, and another 144
possible,
but uncertain white dwarf and hot subdwarf stars from the 1360 deg$^2$
of DR1.  We find 108 of our white dwarf stars are already present in the McCook \& Sion
(1999) catalog. We present a more complete description of the overlap between the two
catalogs in the Appendix.

Another comprehensive spectroscopic survey, the 2dF QSO Redshift Survey \citep{boy00},
has also produced a sizeable catalog of white dwarf stars \citep{ven02}.  The
2dF survey went a bit deeper than the SDSS, but covered a smaller solid angle:
740~deg$^2$ with $18.4 \leq B \leq 21.0$ \citep{ven02}.  Their results include
942 spectroscopically-identified DA white dwarf stars, providing another large
increase in the number of known white dwarf stars.

\section{Object Identification}

The spectroscopic reduction pipeline of the SDSS does not classify stellar
objects with much detail, so we cannot rely on the standard reduction
output to accurately select out white dwarf and subdwarf stars from the 
myriad of stellar (and non-stellar) spectra available.
We thus have to identify  a set candidate white
dwarf spectra, then manually examine each spectrum to determine the
object's classification.

The first task, then, is to identify the candidate spectra.
While there is a specific white dwarf targeting
category, the relative priority of this category is low and we use it
mainly to search for the potentially coolest of the white dwarf stars
\citep{har01}.  We thus cannot view the white dwarf targeting category as
anywhere near a complete sample of candidate white dwarf spectra and therefore 
rely mainly on color and proper motion information to choose our candidates.
The resulting candidate spectra were targeted by a 
variety of SDSS targeting categories.
Table~\ref{tb:objtypes} summarizes the SDSS targeting criteria
that were used to obtain each of our identified white dwarf and hot subdwarf spectra.
Note, this table just lists the targeting category that
was actually responsible for the fiber allocation; it may well be that an
object matched the criteria of multiple target selection categories, but
was ultimately allocated a fiber by only one of them.  \citet{bla03}
and \citet{edr} give more details on the targeting process and provide a
description of the different targeting categories.

The targeting categories listed in Table~\ref{tb:objtypes}: GALAXY, QSO, and STAR\_WHITE\_DWARF
are self-explanatory; ROSAT is for a variety of ROSAT \citep{vog99}
sources as thoroughly described in \citet{and03};
SERENDIPITY has several sub-categories:
MANUAL are targets manually selected and assigned a fiber for any variety of
reasons, DISTANT
looks for unresolved sources that are distant {\it from the stellar locus}:
either very red in $(g-r)$ or very blue in $(r-i)$, and BLUE
is for objects that are particularly
blue in $(u-g)$; the STAR category also has several sub-categories:
BHB are potential blue horizontal branch stars while CATY\_VAR are potential
cataclysmic variables; QA objects are purposefully-selected targets which
have been observed on another plate and whose repeat observations are used
for quality assurance purposes; and finally the HOT\_STD category represents
hot standard
stars used for spectrophotometric calibrations.  As Table~\ref{tb:objtypes}
shows, only a very small percentage of objects were targeted directly as potential
white dwarf stars and most are simply selected as blue objects from the
SERENDIPITY, QSO, and HOT\_STD categories.  \citet{ric02} discuss the SDSS QSO
selection algorithms and describe the overlap in QSO and white dwarf
color-color space in the SDSS.

\setcounter{footnote}{0}

All of the 13 DR1 DOs, however, were targeted by the HOT\_STD algorithm.
Since the HOT\_STD spectra are used for spectrophotometric calibration and the
space density of these objects is relatively low, the category is allocated
fibers with a very high priority.
Thus, the DO,  along with the hot DA and subdwarf, sample from the SDSS should 
be highly complete for isolated stars\footnote{\citet{har03} found 90\%
completeness in a smaller area of the sky for white dwarf stars with \Teff $>$
22000K and $15<g<19$.}, while
the completeness of the rest of the white dwarf subgroups is not as high and
difficult, although theoretically possible, to calculate. The HOT\_STD
algorithm picks isolated stars with dereddened $(u-g)$ and $(g-r)$ colors
between --1.5 and 0 mag and $g$ roughly between 14 and 19 mag.
Since we have SDSS color
information and USNO-derived proper motions \citep{mun03}
for every SDSS-detected photometric object, and we know which ones
were allocated a spectroscopic fiber by what algorithm, and we know which of 
these ended up being which type of white dwarf star, we have a hope of
untangling the complicated selection function and estimating a complete white dwarf luminosity
function in the area of the SDSS.  We leave that project, however,
to future work.

Despite their relative structural and evolutionary simplicity, white dwarf stars are actually quite varied as a
group, and thus we used several different criteria to try to assemble a
list of all SDSS white dwarf candidate spectra.
Because we knew the task of combing through the resultant spectra was going to
be time-consuming, we prepared the list of candidates in the summer of 2002,
before the the final DR1 photometric reductions were complete and before the
final DR1 spectroscopic sample had been settled upon.
Therefore, all photometric selections were made on an earlier
version of the photometric pipeline than what ultimately appeared in DR1 and
our candidate list was different from that which we would get now
were we to repeat the process on the final DR1 data set.
The differences in the photometry due to the
changes in the pipeline are mostly minor and our selection criteria were
purposefully broader than likely necessary, so we expect very few (if any) true
white dwarf stars were missed by photometric criteria.
Since the final
selection of exactly which spectra would appear in DR1 was not ready at the
time of our candidate selection, we sometimes looked at a spectrum
for an object that was ultimately included in DR1 but with a different spectrum
taken at a different time than the one we analyzed.  In these cases,
we have made the cross-assignment of our identification to the new spectrum.
We looked at many of the repeat spectra by eye and found identification changes
only in cases where one spectrum had a significantly higher signal-to-noise
ratio (S/N) than the other.

Table~\ref{tb:selection} summarizes the selection criteria we used to generate
our white dwarf candidate list. The criteria are written with the same
field/quantity names as are available in the SDSS database so the search can be
directly repeated.  These
criteria resulted in 10,800 spectra of 9,400 unique objects.  About two thirds of
these objects were ultimately included in DR1.

Once we had our candidate list, we manually inspected each spectrum and
made a coarse identification.  We then sorted the results into different
categories of white dwarf stars and non-white dwarf stars and fit the DA and DB
spectra with models for \logg and \Teff determinations 
(see below). We did not fit the spectra of the non-DA/DB
white dwarf stars due to their inherent increased 
model complexity. We note that nine of the white dwarf + main sequence binary
systems
reported in \citet{ray03} did not make any of our selection cuts, although we 
have manually added them to our tables for completeness.

Our spectroscopic identifications follow the convention of \citet{sio83}.
Briefly, DA, DB, and DO
stars are white dwarf stars which show lines of H Balmer series, HeI, and HeII
respectively. DQ stars show C lines; DZ stars have metal lines; DH stars
show signs of magnetic fields; and DC stars show continuous spectra, showing no
discernible spectral features.  Hybrid stars are indicated with the dominant
component's symbol first; the secondary, later. For example, a DAB star is one
with dominant H lines, but that shows some HeI lines as well.

\section{Human Identification Complexities}

There were many complications in identifying some classes of white dwarf stars
from our candidate list.
Low S/N spectra (typical for magnitude 20, or greater, objects)
pose a classification problem as noise can make the observed 
hydrogen or helium lines appear broader than they really are.
Where the widths of the lines
were judged to be likely broad, but uncertain, we checked as to whether
a proper motion was detected above about 15 \masy.  If there was such a
motion, we included the star in the DA list; if not, it was called DA:,
the {\it :} suffix indicating an uncertain identification,
and placed in the uncertain white dwarf list.  We also used the $(u-g)$ and 
$(g-r)$ colors 
to help separate the main sequence A and F stars from the DA and sdB stars.
Main sequence A and F stars (and even some horizontal branch 
stars) may appear to have similar line widths as sdB stars but 
very different $(u-g)$ colors (due to the larger Balmer jump).  

Very hot stars pose problems because they generally have weak 
lines and are likely too distant to have a detectable proper motion if they are
white dwarf stars.  These objects might  be classified DA: or sdO: based on 
spectral features judged to be possibly present.

The cool DB stars also pose a special problem.  The line widths of He~I
in these stars become quite narrow and similar to those of the sdO (or He-sdB)
stars.  The latter, however, have substantially stronger He I 4388\AA\ 
relative to He I 4471\AA\ than do the true DBs.  We therefore
relied on this criterion to make the judgment, but if the spectrum was 
poor and line strengths difficult to judge, we used the detection of a 
meaningful proper motion ($> 15$ \masy) as sufficient to classify the star a DB.
Otherwise, we judged it based on the appearance of the low S/N He I 
lines to be either DB: or sdO:.  The colors are, unfortunately,
similar for these two types.  

The proper
identification of featureless spectra can potentially be either a DC white
dwarf or an
extra-galactic object with weak or no features (i.e., a BL~Lac object).
Unfortunately, these two classes can overlap considerably in color.  The designation
was BL~Lac if the source had a counterpart in the FIRST (Becker, White, \&
Helfand 1995) radio or ROSAT \citep{vog99}
X-ray surveys, since cool white dwarf stars are not radio or X-ray sources.
Here again, we used the detection of a  meaningful proper motion as a 
valid basis for choosing the DC classification.  \citet{and03} also discuss
some of the difficulties and techniques in identifying featureless SDSS
spectra.

There were a number of cases where the proper motion and colors 
indicated that the object is a white dwarf, but the spectral type 
is uncertain due to uncertain spectral features seen.  We used 
the identification WDDB:, for example, if possible He I lines were judged to 
be in the spectrum, the {\it WD} indicating a certain white dwarf star and the
{\it DB:} indicating it might be a DB. Likewise, this ambiguity could occur for 
hybrid spectra.  For example, a DB spectrum showing a possible 
but not definite H$\alpha$ line would be denoted DBA:. (Note that the 
Balmer decrement is very steep in DBA stars --- often only H$\alpha$
is detectable.) 
 
\begin{figure}
\figcaption{\label{fig:ugr}$(u-g)_\circ$ vs. $(g-r)_\circ$ color-color diagram 
for the
many types of white dwarf stars identified here.  The WDM classification refers
to white dwarf stars with any non-degenerate companions, virtually always an M
or sdM dwarf star.}
\plotone{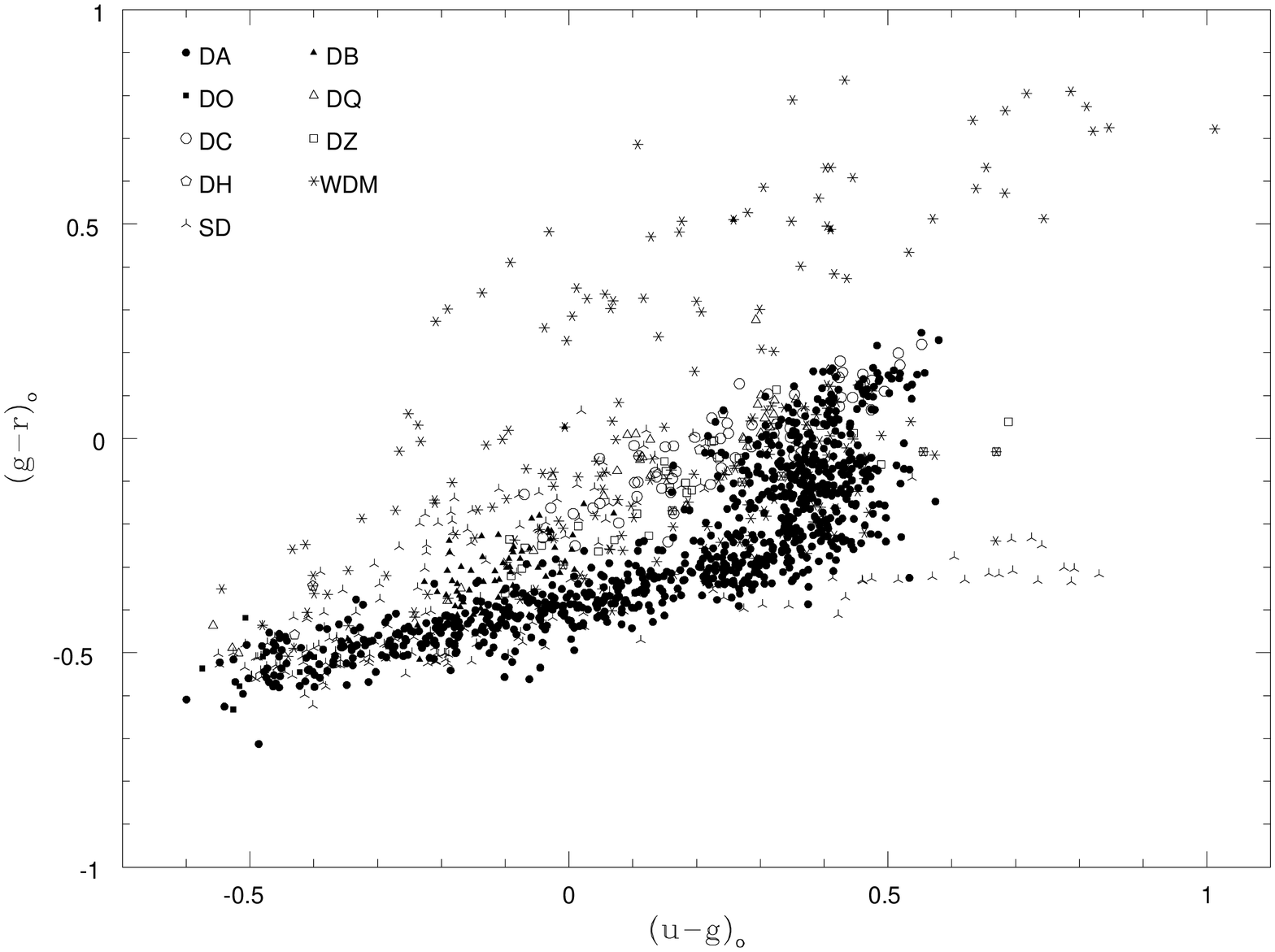}
\end{figure}

\begin{figure}
\figcaption{\label{fig:gri}$(g-r)_\circ$ vs. $(r-i)_\circ$ color-color diagram for the
many types of white dwarf stars identified here.   The WDM classification refers
to white dwarf stars with any non-degenerate companions, virtually always an M
or sdM dwarf star.}
\plotone{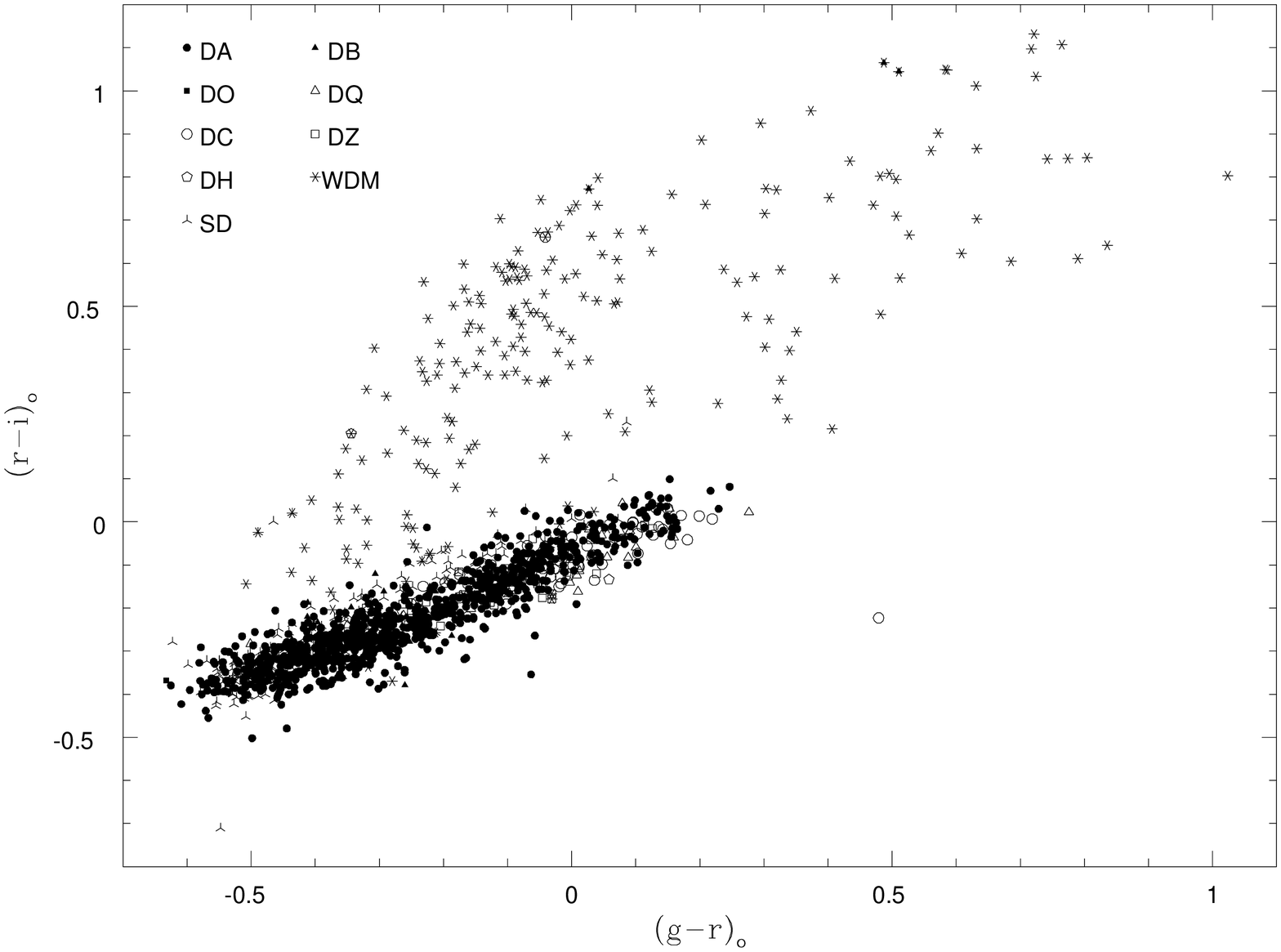}
\end{figure}

Figures~\ref{fig:ugr}~and~\ref{fig:gri} show the resulting $ugr$ and $gri$
SDSS color-color diagrams for identified white dwarf objects.  We did not plot
any objects that have magnitudes that are flagged as bad photometry by the SDSS
pipeline (see discussion below).
The SDSS
photometric system is well described by \citet{edr}, \citet{smi02}, \citet{hog01}, 
\citet{fuk96}, and \citet{gun98}.
The plotted colors are
observed point-spread-function (PSF) magnitudes from the {\it best}
version of the SDSS photometric
database\footnote{The SDSS photometric database includes two sets of data for
each object.  The {\it target} version is the original photometric
detection reduced with whatever the current photometric software was
at the time. This version represents what was used to determine
spectroscopic target selection.  The {\it best} version could be the same,
or later, detection of the object reduced with the latest version of
the photometric pipeline. Except when investigating targeting effects,
the {\it best} sky version is usually the appropriate one to use.},
with full extinction/reddening corrections \citep{sch98} applied to each
object.  Of course, it is not correct to apply the full reddening
correction to every object since some will be close to us and in front of
the dominant extinction sources.  However, it is also not easy to determine
exactly how much extinction is appropriate for each individual object.
To be consistent with \citet{har03}, though, 
we chose to apply the full correction always (and indicate
this in our colors/magnitudes with a subscript $_\circ$).
Since most of our objects are at high galactic latitudes, the extinction is
often small, at any rate.
Our color-color plots clearly show
the separation of the hotter DAs from the more continuum-like DBs and DCs.
The WDM points represent white dwarf stars with non-degenerate companions
\citep[see][]{ray03}.

Some features seen in these plots cannot be interpreted as
indicative of the white dwarf population at large. Since the SDSS spectroscopic
targets are chosen based largely on color cuts in SDSS photometry, there are 
significant selection effects in our sample. However, it is nice to note the
expected continuation of the
identified subdwarf stars into the realm of the blue horizontal branch stars
(between $(u-g)$ of $\approx$ 0.5 to 1.0 ) which demonstrates the arbitrary
separation between the hot subdwarf and extended horizontal branch stars.
These objects have colors which indicate temperatures intermediate between
the traditional horizontal branch and the normal sdB stars.

\section{The Tables}

We present several tables of our spectroscopically identified SDSS white dwarf
and subdwarf samples.  Tables~\ref{tb:DAs} and~\ref{tb:DBs} list the DA and
DB white dwarf stars, respectively, along with their model-fit \Teff and
\logg as described in the next section.  These tables are ordered by fit
temperature.  Table~\ref{tb:all}
lists all the white dwarf and hot subdwarf stars we have identified, including the
human-ID for each one (DA, DB, DQ, etc.).  Finally, Table~\ref{tb:unc} lists
all the objects that we are less certain of, but which could be white dwarf or hot
subdwarf stars. The latter two tables are ordered by right ascension and
declination.  All our tables are also online at the SDSS DR1
value added catalog site: 
{\url http:///www.sdss.org/dr1/products/value\_added/wdcat/dr1/index.html}.
The online catalogs have links to the SDSS spectra and finder charts, as well
as to plots detailing our model fits.

We do not separate the DCs, DQs, DHs, DZs, or binary white dwarf stars in these
tables, but refer the interested reader to \citet{har03} for a more general
discussion of these other white dwarf subtypes, \citet{sch03} for a
discussion of the magnetic white dwarf stars, \citet{lie03a} for a discussion of
white dwarf stars with carbon and oxygen lines, \citet{ray03} for a discussion
of white dwarf plus main sequence M star binaries, and \citet{krz04} for a
discussion of the SDSS DO white dwarf sample. Finally, \citet{muk04}
describe the discovery of 32 new DAV (ZZ~Ceti) pulsating white dwarf stars
from our DA sample and include the many SDSS DAV candidates observed that were
ultimately non-pulsators, and \citet{nit04} discuss new DBV (V777~Her)
pulsators resulting from the new DR1 DBs.

Each entry in the tables starts with the object's official SDSS name.
The format of the name is SDSS~JHHMMSS.ss+DDMMSS.s 
where the HHMMSS.ss is the truncated (not rounded)
hours (HH),  minutes (MM), and seconds (SS.ss) from the SDSS J2000 right 
ascension
and the +DDMMSS.s is the truncated declination in degrees, minutes, and
seconds. Each object should be referred to by these names.  We also  provide
the right ascension and declination in decimal degrees, as that is a convenient
format for use in SDSS databases.  SDSS astrometry is thoroughly discussed in
\citet{pie03} and reported positions are good to less than 0.1'' rms.
Since the SDSS coordinates are all reported for equinox 2000.0 at the epoch of
the {\it best} imaging observations, we also include this epoch in our
general catalogs, Tables~\ref{tb:all} and~\ref{tb:unc}.
In addition, we list the SDSS plate, MJD, and fiberID, needed to uniquely
obtain the SDSS spectrum from the DR1 data archive server
(\url{http://das.sdss.org/DR1-cgi-bin/DAS}).  The plate number describes the
pre-drilled plate of 640 fibers with which each object was observed. The
fiberID details which of the 640 fibers gathered each particular spectrum and
the MJD is the SDSS-modified Julian date on which the data were taken.
Next are the $ugriz$ PSF magnitudes and uncertainties.
along with the final median S/N squared per pixel of the spectrum from the DR1
{\it best} database.  We precede each magnitude that is flagged by the SDSS
pipeline as questionable with an {\it *}.  Table~\ref{tb:badflags} lists the
photometric flags we checked along with a brief description of each flag.
See \citet{edr} and \citet{dr1} for more details of all
SDSS outputs.

The DR1 database contains a proper motion, where measured, for each photometric
object, but the proper motions we present here are not those from the SDSS
database.  Instead,
proper motions have been measured with the following procedure.
First, we matched each of our objects against the USNO--B1.0 
catalog \citep{mon03} by finding objects in USNO--B within 1''
of each SDSS position at the epoch of the SDSS imaging scan.
Next, we extracted the position of each detection of the matching object
from USNO--B, giving up to five measurements and
epochs on the five sky survey plates included in USNO--B.
We corrected each of these USNO--B positions for systematic errors by
subtracting the mean difference between the SDSS and USNO--B positions for the
nearest 100 galaxies in the
magnitude range $17 < g < 19.5$. 
Finally, we recalculated the proper motion using the SDSS position
plus all (up to five) USNO--B positions.  We used a weighted linear
solution, assuming errors of 45 mas for the SDSS
position
and 120 mas for each sky survey position in each coordinate
for determining the relative weights in the solution.
This procedure improves proper motions by a significant amount and
is described more fully by \citet{mun03}.
The motions have rms errors of 3--5 \masy in each coordinate.
We generally consider any proper motions less than 12 \masy as not
significant.

Finally, each table also has the extinction in the SDSS {\it g} filter, A$_g$,
as stored
in the SDSS photometric database.  These values use the reddening maps as
presented in \citet{sch98}.  Table~\ref{tb:reddening} lists the multiplicative
factors for converting $A_g$ into the extinctions in the other SDSS filters, as described by
\citet{sch98} and \citet{edr}.  We also
include various entries in the {\it Notes} column to highlight certain objects
or indicate the results of our human checks of the computer fits (described
below).

The DA and DB tables also list our model-fit-determined 
\Teff and \logg and their corresponding uncertainties along with the $\chi^2$
of the best fit.   Following the \Teff
and \logg determinations is the ASCII ID for each spectrum from our fitting
code.  The format is simply DA or DB followed by the \Teff subtype
(50400K/\Teffns) and the \loggns, separated by an underscore.  The {\it CH:} prefix
and colon suffix are quality assessment checks which are discussed in the next
section.  Where our check on the auto fitting deemed the fit invalid, we
replace the fit parameters with zeroes and the fit ID with {\it N/A}.
The combined and uncertain white
dwarf tables list only the human ID for each object, since we only fit models
to the DA and DB stars.  A colon in the human ID
means the identification is uncertain, but only the modifier immediately
before the colon is deemed uncertain.  For example, DA9: refers to a star that
we consider to be a DA, but which we think is a DA9 subtype.  (Most of our
human IDs, however, do not include subtypes.)

\section{Model Fitting}

We fit each human-identified DA and DB spectrum to theoretical models from
Detlev Koester to
determine their  temperatures and surface gravities.  
Koester's  models are
described in Finley, Koester, \& Basri (1997), although we received an
updated grid at the time we started our analysis. One change in the current
models from those described in \cite{fin97} is the use of the now-standard
ML2/$\alpha=0.6$ convection model. The hydrogen
atmosphere models range in \Teff from 6000 to 100,000 K and in \logg
from 5 to 9.  The helium atmospheres range in \Teff from 10,000
to 40,000 K and in \logg from 7 to 9.  Our procedure is to measure
the $\chi^2$ statistic on the difference between the observed spectra
and the models, using the quoted errors from the SDSS spectroscopic
pipeline.  For DA stars, we use only the wavelength range 3870\AA\ to 7000\AA;
for DB stars, we use 3870\AA\ to 5400\AA (for reasons described below).
We exclude pixels that are flagged by the spectroscopic
pipeline in the ``AND'' mask (meaning that the pixel was masked in 
all exposures) with the bits {\tt 0x1fff0000}.  This mask includes all
single-pixel failure modes.  In addition, we visually inspected all
cases in which one or both cameras of the spectrograph were flagged
with the full-chip mask bits {\tt 0x1cf}.  The results of the visual
inspection are indicated in the {\it Notes} section of the tables,
with ``1'' and ``2'' meaning trustworthy and untrustworthy, respectively.

The SDSS spectrographs have a typical instrumental dispersion of about
170 \kms\ FWHM.  Because this is considerably less than the typical line widths
in white dwarf stars, smoothing the models to the instrumental resolution (or
not) does not affect the fits much.  We have not smoothed the models for
the fits presented.  The error introduced by this for DAs is less than
1\% in temperature and generally negligible (0.02 dex) in \loggns; however,
at temperature below $10^4$K (where our input model grid has systematic
errors anyway), our gravities are biased high by 0.1--0.2 dex.  For DBs,
the effects on temperatures are small (less than 3\%), but the gravities
are biased high by 0.1 dex for temperatures below 20,000K.  We plan to
include the fits with instrumental dispersion in future samples which will be
available at the SDSS value-added catalog URL given earlier.

Because of unknown reddening and the desire to insulate the procedure
from spectrophotometry errors, we permit the fit the freedom to re-flux
the models according to a low-order polynomial.  This is done efficiently
by performing the $\chi^2$ fit as a linear least squares optimization
to a set of vectors defined by the model spectrum and the model spectrum
multiplied by a series of smooth basis functions.  We use the first seven
Chebychev polynomials (with the first being a constant) in linear wavelength
as our refluxing basis.  By using
the minimum in $\chi^2$ for this 7-dimensional optimization as
our basis for comparison between models, we are effectively marginalizing
over the refluxing parameters. 
The shape of the Cardelli, Clayton, 
\&  Mathis (1989) extinction curve over this octave in wavelength is 
extremely well modeled by a 6th-order polynomial, and so our procedure
fully marginalizes over reddening uncertainties.

Due to the near degeneracies in the line strengths and profiles
of white dwarf stars,
we supplement the spectroscopic fitting with additional information from
the SDSS photometry.  Each model is convolved with the SDSS filter curves
to yield predicted colors.  The SDSS photometric zeropoints nearly but not 
exactly satisfy the AB convention \citep[see][]{edr,dr1}.  
We correct the photometric zeropoints from
the AB system to the SDSS system by $u(AB)=u(SDSS)-0.04$ and 
$-0.01$, $0$, $0.015$, and $0.03$ for $g$, $r$, $i$, and $z$, 
respectively.  These corrections are approximate and still subject
to change.
We then construct the $\chi^2$ statistic
for the difference between the observed colors and the predicted colors
using the quoted errors in the five bands, with a systematic floor of 0.007
mag added in quadrature (0.015 mag in $u$ and $0.010$ mag in $z$).  
To account for reddening, we apply a baseline
correction of 50\% of the \citet{sch98} reddening map and then marginalize
over the reddening direction in color space (assuming $R_V=3.1$) with a
1--$\sigma$ prior of 50\% of the predicted reddening.  
We then forbid reddening values less than zero and penalize
values above the \citet{sch98} value with a prior of $90\%\pm10\%$ of the
predicted value (thereby yielding a continuous $\chi^2$ distribution).
In other words, we adopt a reddening prior of $0.5\pm0.5$ of the predicted 
value with a steeper wall at unity and a sharp cutoff at zero.  
Clearly, this is an approximation, but our primary goal is to pick the 
correct local minimum in degenerate cases.
We sum the spectroscopic and photometric $\chi^2$ with equal weight.

Having computed the $\chi^2$ for each model, we estimate the best fit
and errors in two ways:  1) Use the lowest $\chi^2$ model
with a quadratic fit to the neighboring points to find the interpolated
minimum and errors, and 2)  Convert the $\chi^2$
into a Gaussian likelihood and simply find the moments of this distribution,
assuming a uniform metric in inverse temperature and \loggns.
The fit values presented in the tables are from the likelihood moments.  

In the case of multiple near-degenerate minima, the likelihood estimate
will try to split the difference (and greatly increase the errors), 
while the global minimum will make a somewhat arbitrary choice.  
Hence, we monitor the difference between these two estimates as 
a way to find cases with multiple minima.  
Inspection of near-degenerate cases suggests
that the likelihood error range includes both minima, but the
reader should be aware that the quoted mean fit isn't necessarily
near one of the minima.

Our fitting procedure relies upon a somewhat coarse set of models
between which we interpolate a fine grid.  We have identified two
negative consequences of this procedure.  First, models that fit to
gravities near the edge of the grid, in particular the upper bound
at $\log g =9$, are incorrectly thrown against the boundary and given
very small errors.  This anomaly sets in at $\log g\gtrsim8.8$, and
any quoted numbers above this should be viewed only as indications
of high gravity rather than an accurate fit.
Second, for DA stars of very high signal-to-noise ratio, our model grid
is simply not fine enough.  Essentially we are estimating the quadratic
shape of the $\chi^2$ distribution from points far from the minimum.
We expect that this leads to some lack of accuracy and an underestimate
of errors.  This problem seems to occur when the errors in temperature
fall below 1\% of the temperature itself.  Future versions of our
model fitting will attempt to address these problems.

Since the release of DR1 \citep{dr1}, Tremonti et al. (2003) have made
significant improvements to the spectrophotometry of the SDSS pipeline. These
improvements will be available in the next public data release, DR2, but
having them at our disposal now, we have used these spectra for our fits 
rather than the as-released DR1 versions.  Our tests (described below) of
internal errors of repeat spectra reduced with both pipelines, however,
give us confidence that our results are still representative of the
publicly-available DR1 spectra.

The fitting code performs two levels of checks to monitor the quality of
its fits.  We mark the most severe problems with a {\it CH:} (which stands
for ``Check'') at the start of the ASCII identification output of the code
(included in the DA and DB tables).  The {\it CH:} indicates
that the code could not find any satisfactory fit, but it reports the best
it had.  These checks can be triggered for several reasons: the spectrum
had too many masked pixels, the reduced $\chi^2$ of the best fit was too
high or too similar between the DA and DB best fits, there were two minima
in the likelihood fit surface, the S/N of the fit spectrum was too low,
and/or the photometric colors disagreed significantly from the best fit.
We looked at each of the {\it CH:} cases by eye and determined whether or
not the fit was reasonable despite the program's complaints. Where the fit
was found reasonable, we include fit parameters and a {\it 1} in the {\it
Notes} column; where not, we do not include any of the fit parameters and add a
{\it 2} in the {\it Notes} column.

Other, less severe trouble indicators are marked with a {\it :} appended to
the program's ASCII identification.  In general, however, we find no evidence
that the identifications flagged with colons are any worse relative to their
quoted errors than are the non-colon objects.  The reasons for triggering
these warnings are similar to those for the {\it CH:}, but less severe:
the reduced $\chi^2$ was marginal (between 1.2 and 1.5) or too good $\chi^2$
(indicating perhaps a loss of signal in the spectrum), the $\chi^2$ of both
the best DA and DB fits were marginally similar, the S/N of the spectrum was
moderately low, a DB fit \logg was pegged to an endpoint of the DB grid,
the colors disagreed moderately with those predicted from the best fit,
the necessary refluxing was too large, and/or an improbable reddening value
was needed.

\begin{figure}
\figcaption{\label{fig:fit.432-378}Our model fit to the $g=16.9$ DA, 
SDSS~J074041.67+412107.4.}
\plotone{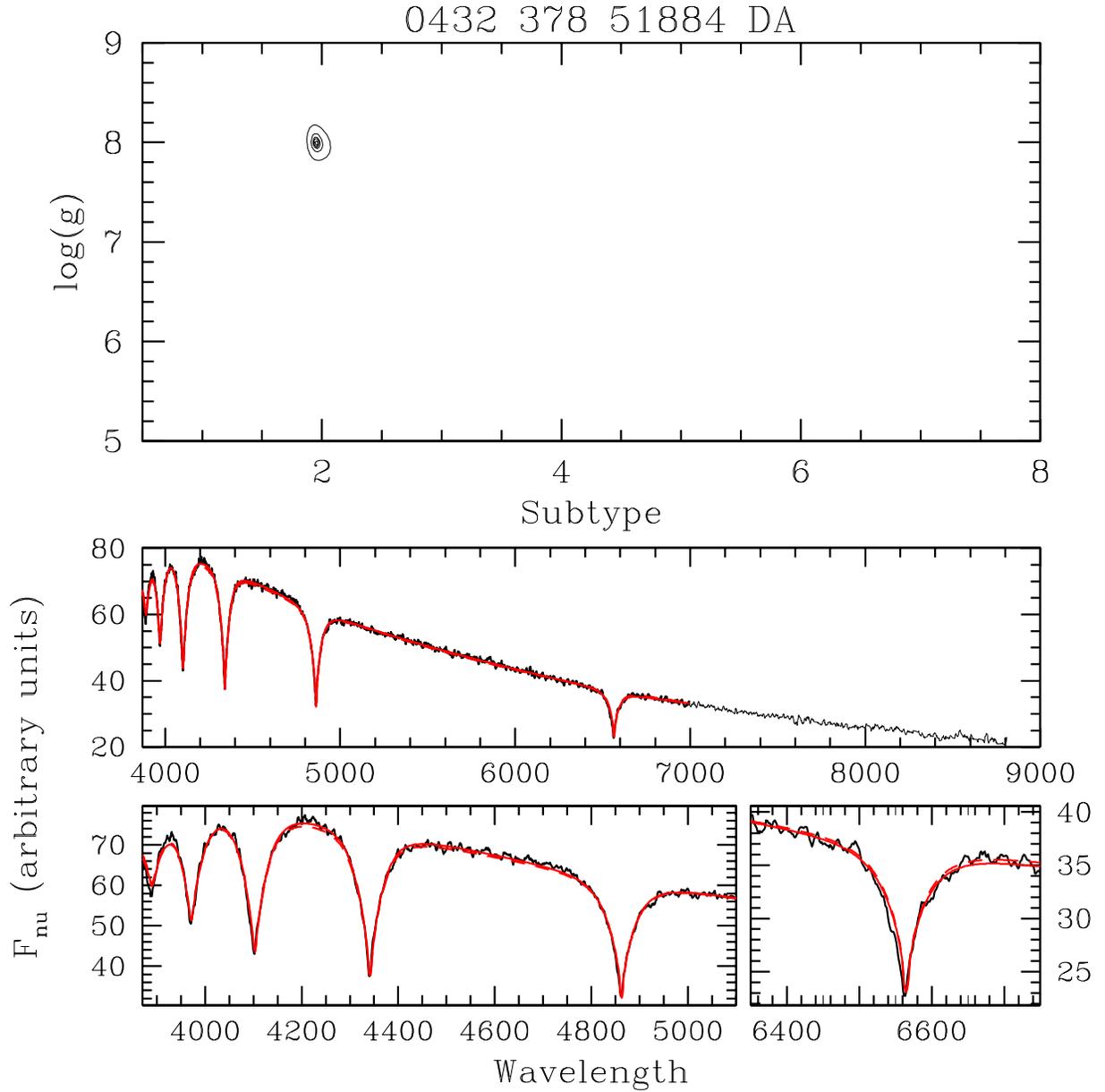}
\end{figure}

\begin{figure}
\figcaption{\label{fig:fit.431-512}Our model fit to the $g=20.5$ DA, 
SDSS~J073651.84+375545.1.}
\plotone{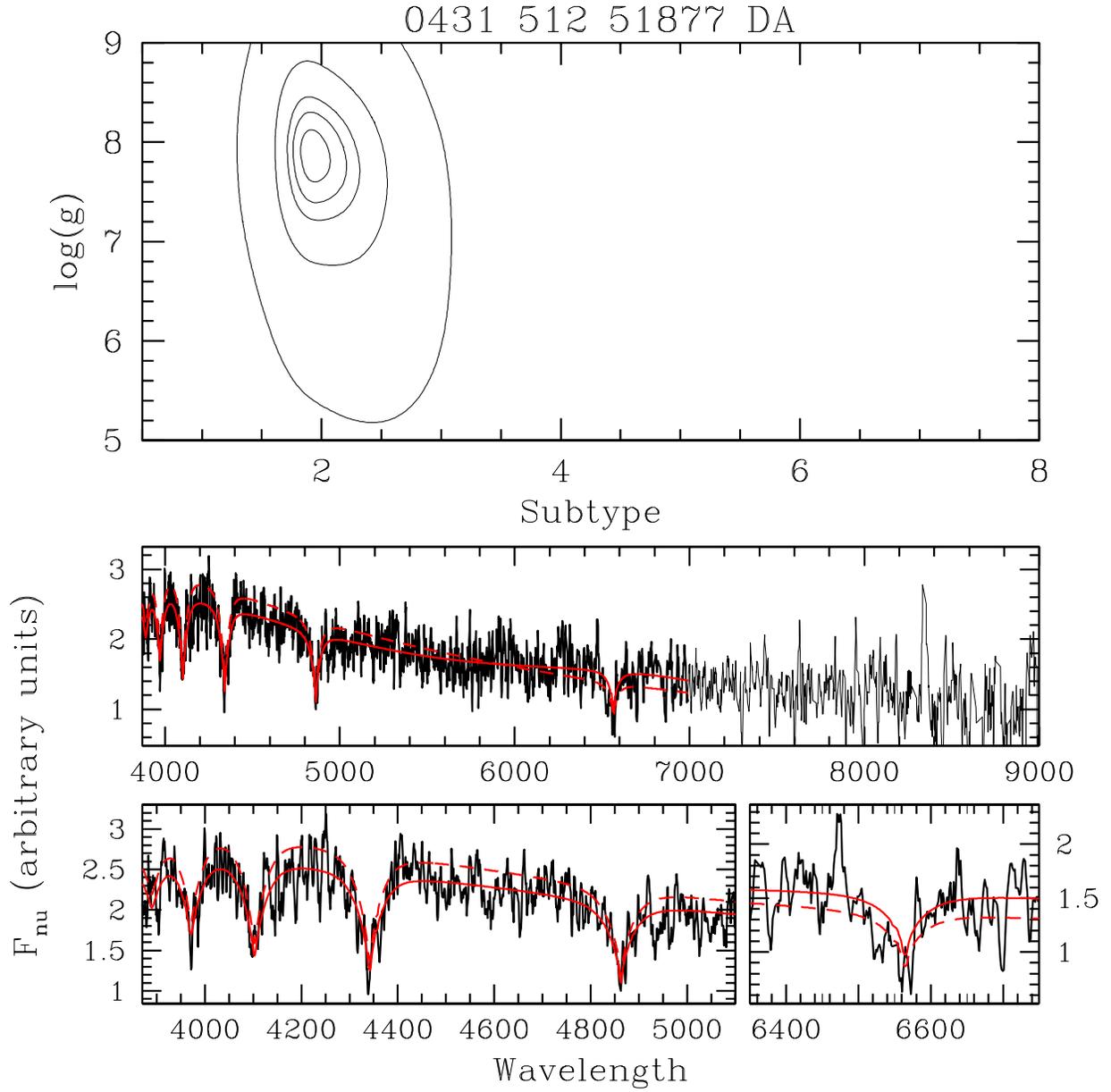}
\end{figure}

\begin{figure}
\figcaption{\label{fig:fit.391-378}Our model fit to the $g=18.9$ DB, 
SDSS~002633.89+005425.9.}
\plotone{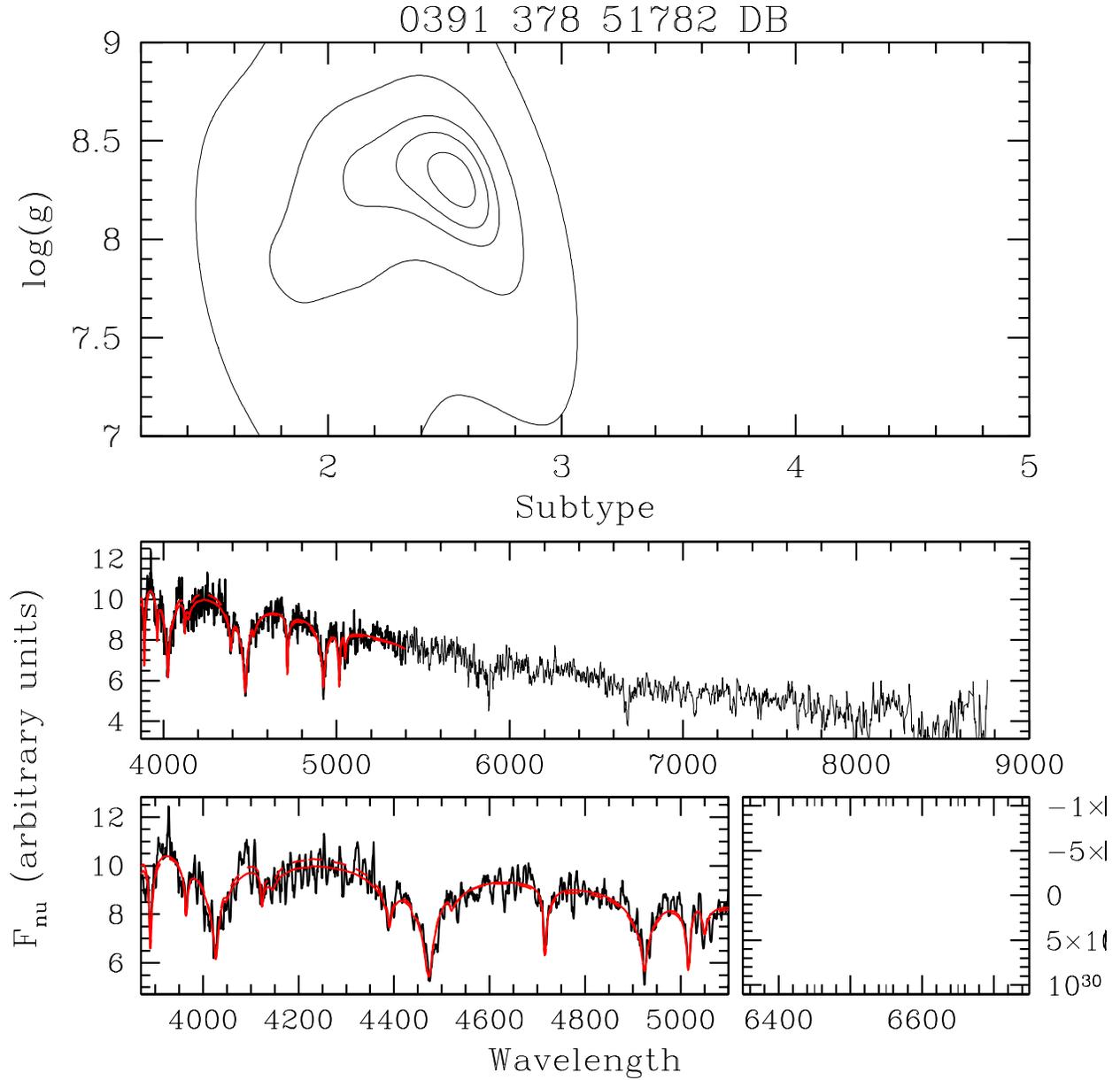}
\end{figure}

Two sample fit DA output plots are shown in
Figures~\ref{fig:fit.432-378}~and~\ref{fig:fit.431-512}.  The fit parameters
are similar for the two objects, but the first has a {\it g} magnitude of 16.9,
while the second object is ${\it g}=20.5$.  Figure~\ref{fig:fit.391-378} is
for a ${\it g}=18.9$ DB. The top half of these plots shows the fit contours
in \logg -- subtype space, with subtype being the usual \Teff expressed in
units of 50400K/\Teff \citep{sio83}. The likelihood contours are at 1, 2, 3,
5, and 10-$\sigma$ (i.e. $\Delta\chi^2 = 2.3$, 6.2, 11.8, 29, and 105 for the
two-dimensional distribution).  The bottom panels show various zooms of the
fit plotted with the spectrum itself (reddened as observed).  The dashed
line fit is the fit without the refluxing (and without any reddening);
the solid line fit is the adopted refluxed fit.

Visual inspection suggests that the model fitting procedure works well for
DAs.  The inclusion of the photometry usually breaks the line-degeneracies
between hot and cold DAs.  We find that at temperatures below $10,000$ K,
the fitted gravities significantly exceed $\log g=8$.  We believe this may be
a failing of the input models (see next section), although in the
extreme limit of $\log g=9$, our procedures encounter the systematic problem
listed above.  We have made no attempt to treat DA+M binaries or DAe emission
line contamination properly, so fits to such objects should not be trusted.
Binaries are usually found because of their severe photometric residuals
in the $z$ band and the absorption bands in the red part of the spectrum.
Subdwarf B stars are generally recovered as low gravity objects, but we have
not attempted to fit the diversity of hydrogen and helium lines that are
found in these stars. Thus, we only present fitted \Teff and \logg values
for what we think are single DA and DB stars.

DB white dwarf stars are more difficult to fit because above about 18,000K,
the line strengths depend more weakly on temperature.  Small systematic errors
in the observations or the models appear to cause large shifts in the fitted
temperatures that significantly exceed the formal errors.  The photometry
provides an additional source of information on the temperatures, but this
can sometimes be ambiguous because temperature and reddening have similar
effects on the colors.  We note that the prior on the reddening is tied to
the predicted extinction, and therefore stars in high extinction regions
will be less constrained by the photometry.  If this leads to a bias, the
magnitude of that bias would be galactic-latitude dependent.  In short,
while the fitting pipeline appears to be correctly finding the hot DBs, we
do not believe that our detailed fits to the hot DBs are reliable.  We have,
however, begun a program to look at the hotter DBs for variability and to
date, have found two out of four well-observed candidates to be variables
\citep{nit04}, suggesting our hot DB fits are at least indicative.

We found that the fitted gravities in cooler DBs tend to be substantially
higher than the conventional value of $\log g=8$.  However, we find that
this result is quite sensitive to whether the lines between 5400\AA\
and 7000\AA\ are included in the fit.  This may indicate subtle errors
in the models, at least in certain lines.  We have not addressed this
source of systematic error save by choosing to restrict our DB fits to
the spectrum shortward of 5400\AA.

The values of the spectroscopic $\chi^2$ are often quite close to one per
degree of freedom even in cases of good S/N.  This is very encouraging, as
it indicates the quoted errors from the spectroscopic pipeline do properly
represent the pixel-to-pixel noise in the spectra.  However, the errors we
find from the standard $\chi^2$ methods are rather small.  To investigate the
validity of our errors, we found 265 of our white dwarf spectra (242 DAs,
23 DBs) that had repeat spectra taken by the SDSS and which our fitting
avoided ``Checks'' on both fits.  In all cases, these repeat spectra are
separate exposures taken on different  nights than the first observation.  Most are simply repluggings of
the same plate (meaning the relation between optical fibers and plug holes
have been permuted) and hence share the same calibration stars but typically
illuminate different detector pixels.  We fit each of the repeat spectra
and compared with the results from the original spectra fits, using the
same photometry in both cases.  In detail, the ``second-epoch'' spectra we
used in this comparison were reduced with the DR1 pipeline, rather than the
DR2 pipeline that we used for our primary set of stars.  The good agreement
we describe below demonstrates that the fits are reasonably insensitive to
continuum changes and fluxing errors.

\begin{figure}
\figcaption{\label{fig:tgnu}The distribution of errors in our fits of repeat
spectra for DAs (left) and DBs (right). The y-axis is
the difference in the \Teff determinations from each fit divided by
the sum in quadrature of the two \Teff uncertainties.  The x-axis is the
similar quantity for \loggns.  The ellipses
are the contours that should include 68\%, 95\%, and 99.7\% of the points if
the distributions were Gaussian (and independent).}
\plottwo{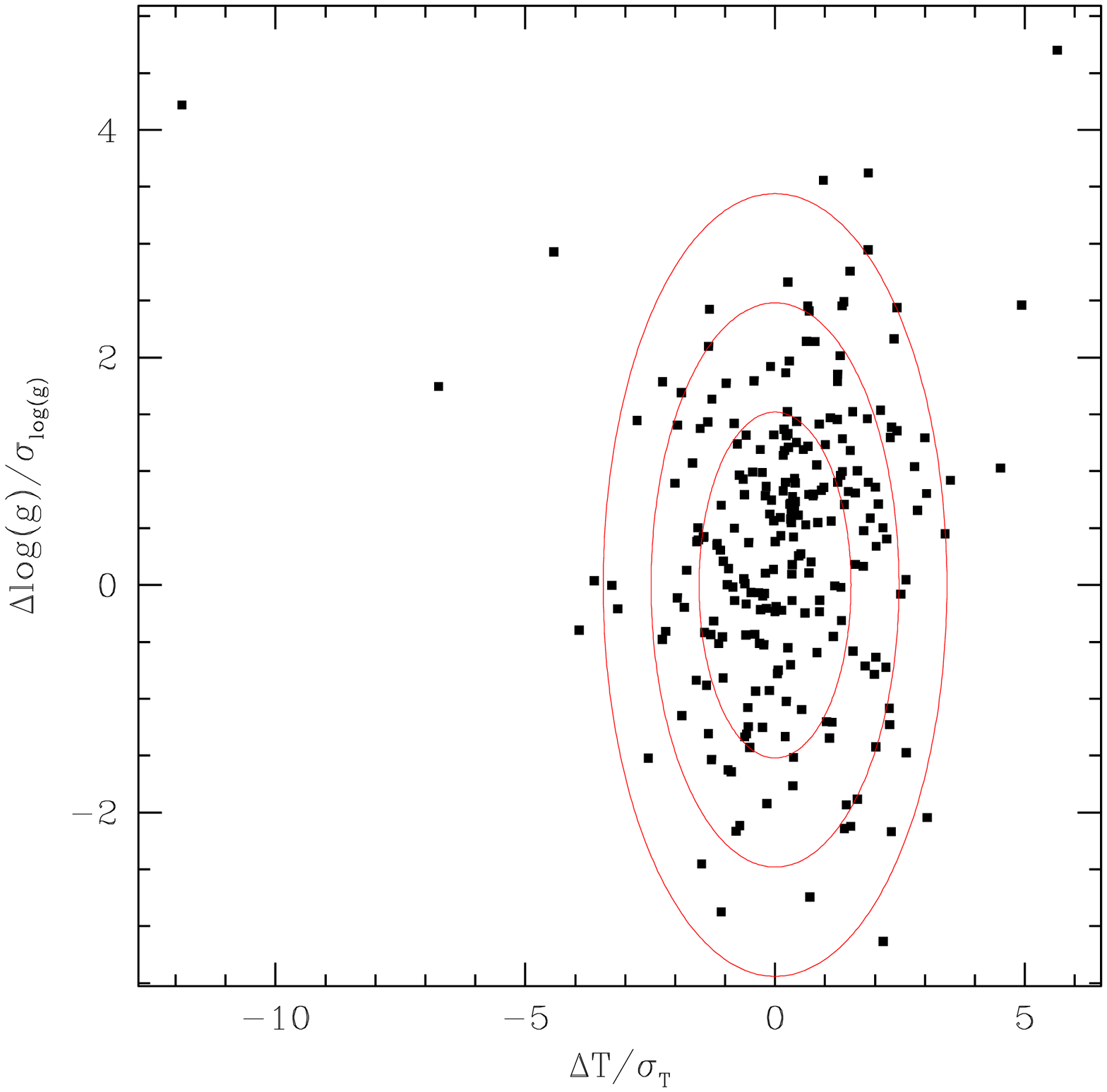}{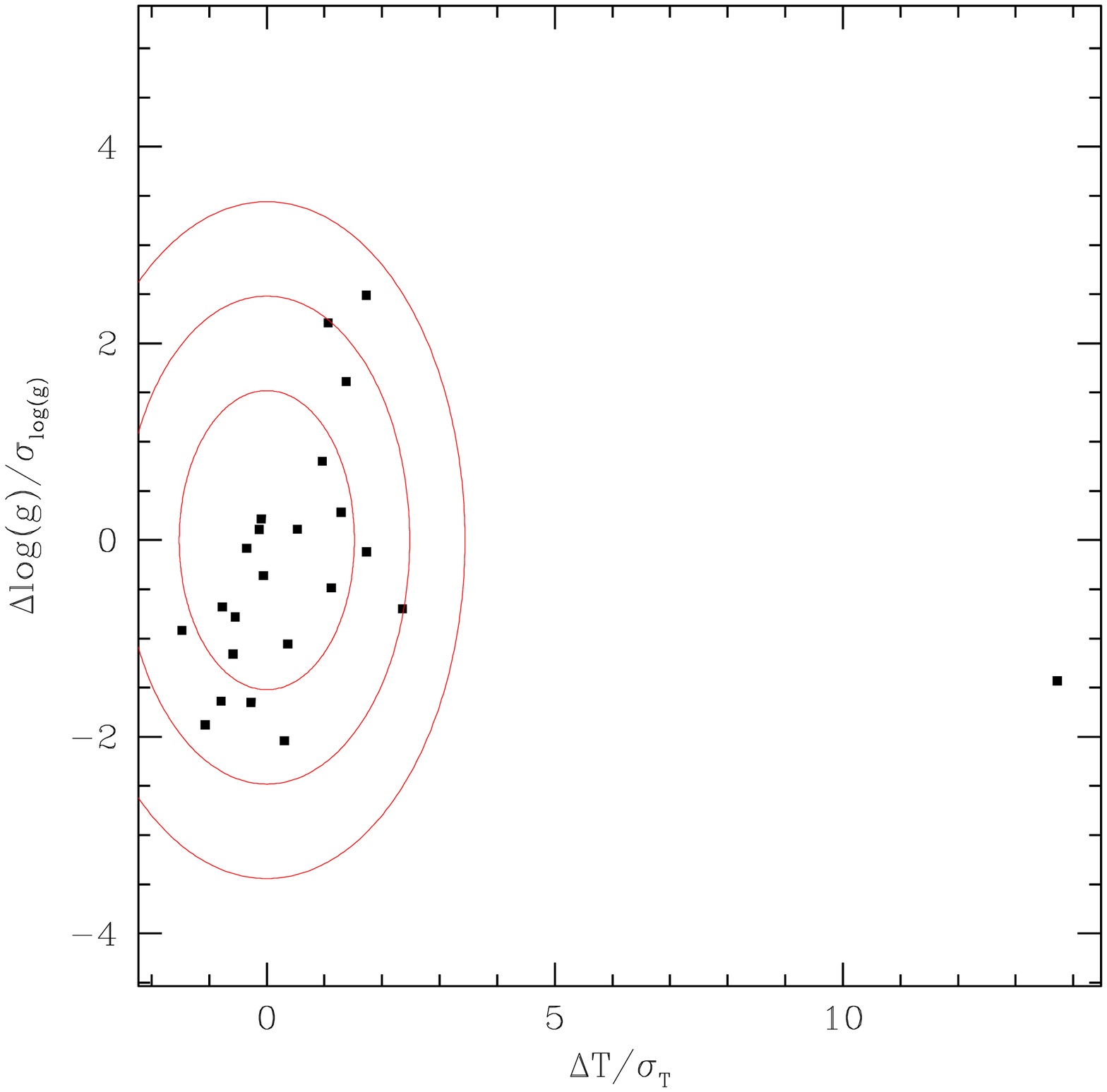}
\end{figure}

Figure~\ref{fig:tgnu} shows the difference between the fitted temperatures
divided by the quadrature summed uncertainties versus the same quantity for \logg
for the repeat DA and DB samples.
The ellipses shown in these plots are the contours that should
include 68\%, 95\%, and 99.7\% of the points if the distributions were
Gaussian and independent.  The DA fit residual distribution is about 40\% larger
than the Gaussian predictions from the formal errors.
There is some hint that the
scatter in those stars where the formal errors are
below 1\% in temperature is slightly worse than the other stars.  However,
it is important to stress that 78\% of the DAs have temperature differences
less than 5\%, and only 5 of the 242 changed temperature by more than 20\%
(at least 3 of these cases are jumps across the temperature degeneracy).
The DB distribution is only about 10\% larger than the Gaussian prediction, 
after excluding the one catastrophic outlier (in which a star jumped
across the temperature degeneracy).
We suspect, however, that this relative agreement is forced upon us by the photometric
weighting in our fits and that our fitting probably is not as good as with the
DAs.  Clearly, systematic errors in the fitting procedure or in the theoretical
models are not tested by the repeat spectra.  In summary, though, we suspect
our DA uncertainties are underestimated by roughly 40\%, and probably a similar
amount for the DB fits.

\begin{figure}
\figcaption{\label{fig:tgavofits}Relative temperature (left) and absolute \logg
(right) differences between our (autofit)
fits and literature (lit) fits for already known DA stars. 
The squares are based on
published fits based on Koester models, the circles are based on Bergeron
models, and the asterisks are based on other models.}
\plottwo{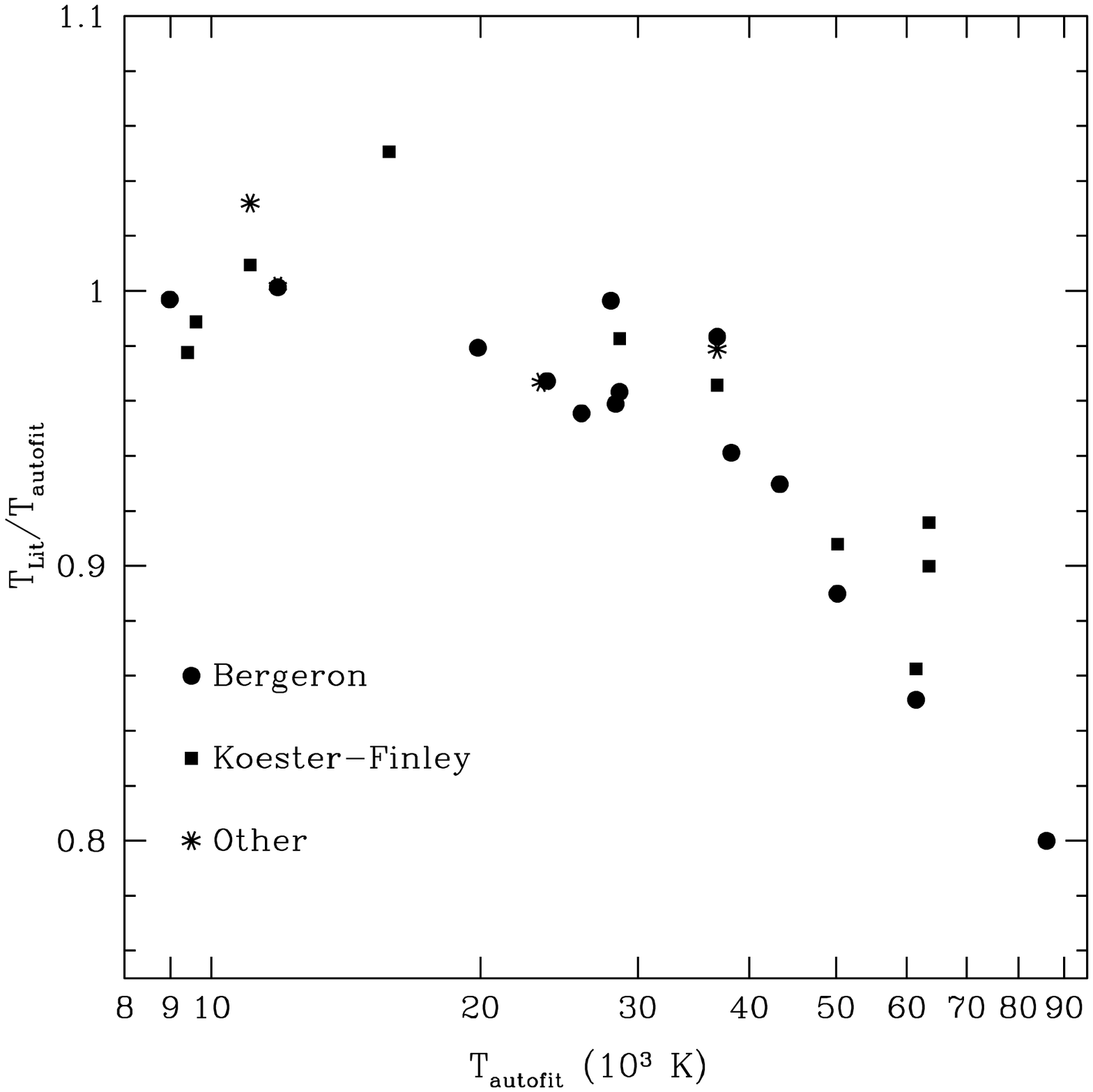}{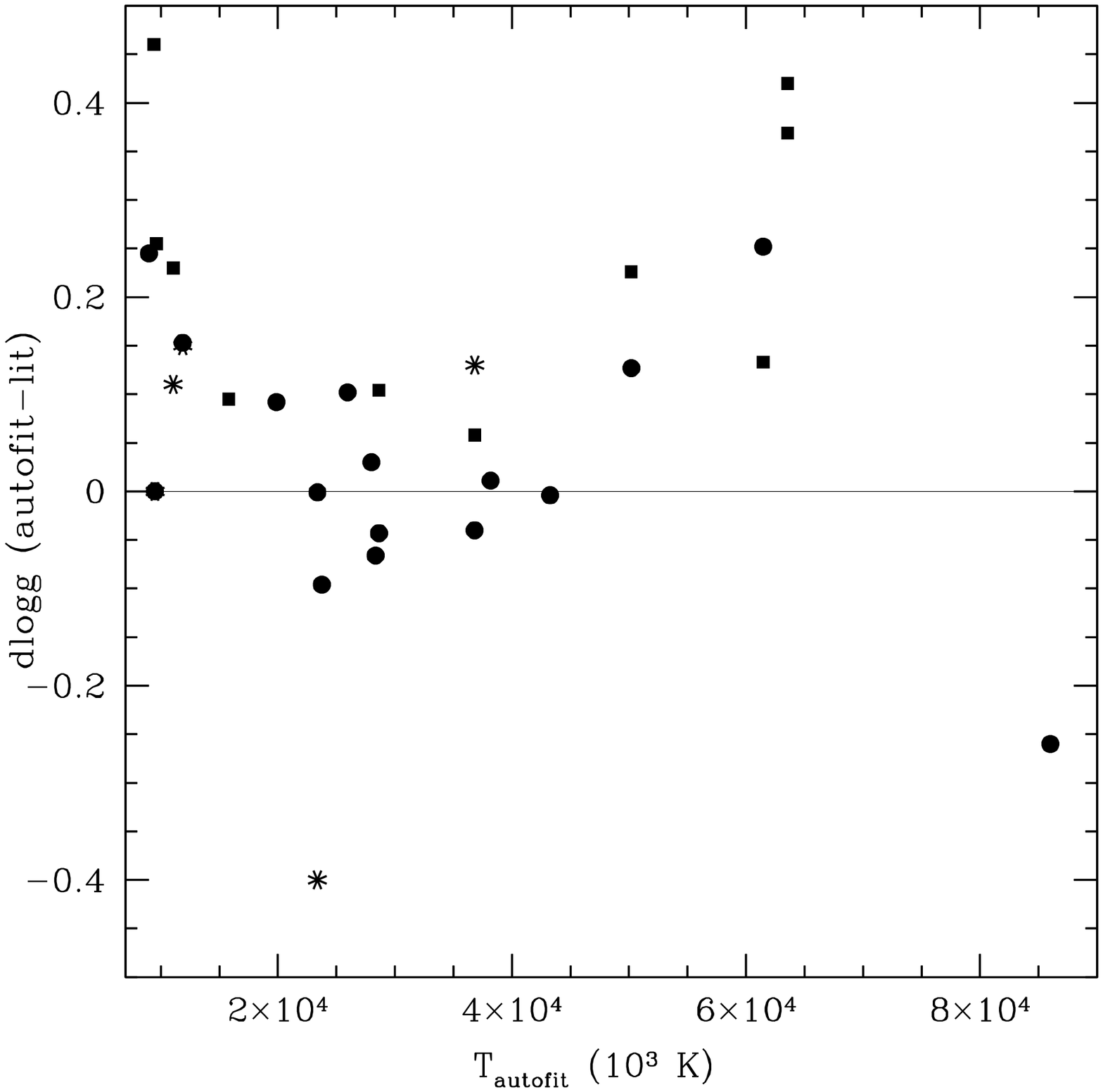}
\end{figure}

Figure~\ref{fig:tgavofits} show the relative \Teff and absolute \logg 
differences
produced by our fitting routines versus those available in the literature for
previously known DA stars. There are only 20 stars used in this comparison,
but some have multiple literature fits which are all included.
The solid squares represent literature
fits based on Koester models \citep{koe01,hom98,fin97}, the circles are based
on Bergeron models \citep{ber92a,ber94}, and the
asterisks are based on other models \citep{mar97,ven97,nap99}.  The figure 
shows a good agreement for \Teff $<$ 25,000K or so, with a 
systematic trend in our fits to overestimate \Teff when
compared with literature values, increasing as \Teff increases to
within 5\% for \Teff $<$30,000, within 10\% for \Teff $<$ 60,000K, and
increasing to as much as a 20\% relative
overestimate for our highest \Teff fit near 90,000K.  The Bergeron and Koester
literature fits agree with each other quite well, so our differences are
probably due to our method of fitting, perhaps our use of refluxed
continuum fits versus the traditional line profile fitting.

The \logg plot
shows reasonably good agreement in the \Teff range around 15,000 -- 50,000K,
but we tend again to overestimate (as much as 0.4) the quantity for stars
much cooler and much hotter than that when compared to the literature fits.
We discuss below  possible problems to the model fits to cool stars,
but our systematic increase in \logg appears to be worse than that found by
other investigators, although the sample of cool stars is small here.
The general trend to both higher \Teff and \logg with increasing
\Teff makes some sense in that our  code is trying to compensate for the higher
\Teff by also increasing \logg (or vice versa).  

We regard this method of model fitting as reasonably accurate, but
we cannot rule out the possibility of as much as 10\% biases in the temperature
scale as discussed above.  The recent successes of \citet{muk04} in finding DAV pulsators
and \citet{nit04} in finding DBV pulsators
based on our fit-determined parameters, and the theory that each instability
strip is purely a function of \logg and \Teffns, give us additional
confidence the fits are reasonable. Indeed, our \Teff fits agree to within 5\%
to literature values for DAs with \Teff $<$ 45,000K.  Like all automated 
pipelines, outliers
should be viewed with some suspicion; the fit and spectra of interesting
cases should be checked by eye before investing telescope time on their further
study. To facilitate these checks, we have made plots similar 
to Figures~\ref{fig:fit.432-378} through~\ref{fig:fit.391-378} available
for all our DA and DB fits in the HTML version of our catalogs, available on
the SDSS DR1 value added catalog web page at
{\url http:///www.sdss.org/dr1/products/value\_added/wdcat/dr1/index.html}.

\section{Discussion}

\begin{figure}
\figcaption{\label{fig:tgood}\Teff vs. $(u-i)_\circ$ from our DA and DB model 
fits.  We include only data with the spectroscopic S/N in {\it g}$>10/$pixel. 
The solid line represents \citet{ber95} models convolved with SDSS filter
curves.}
\plotone{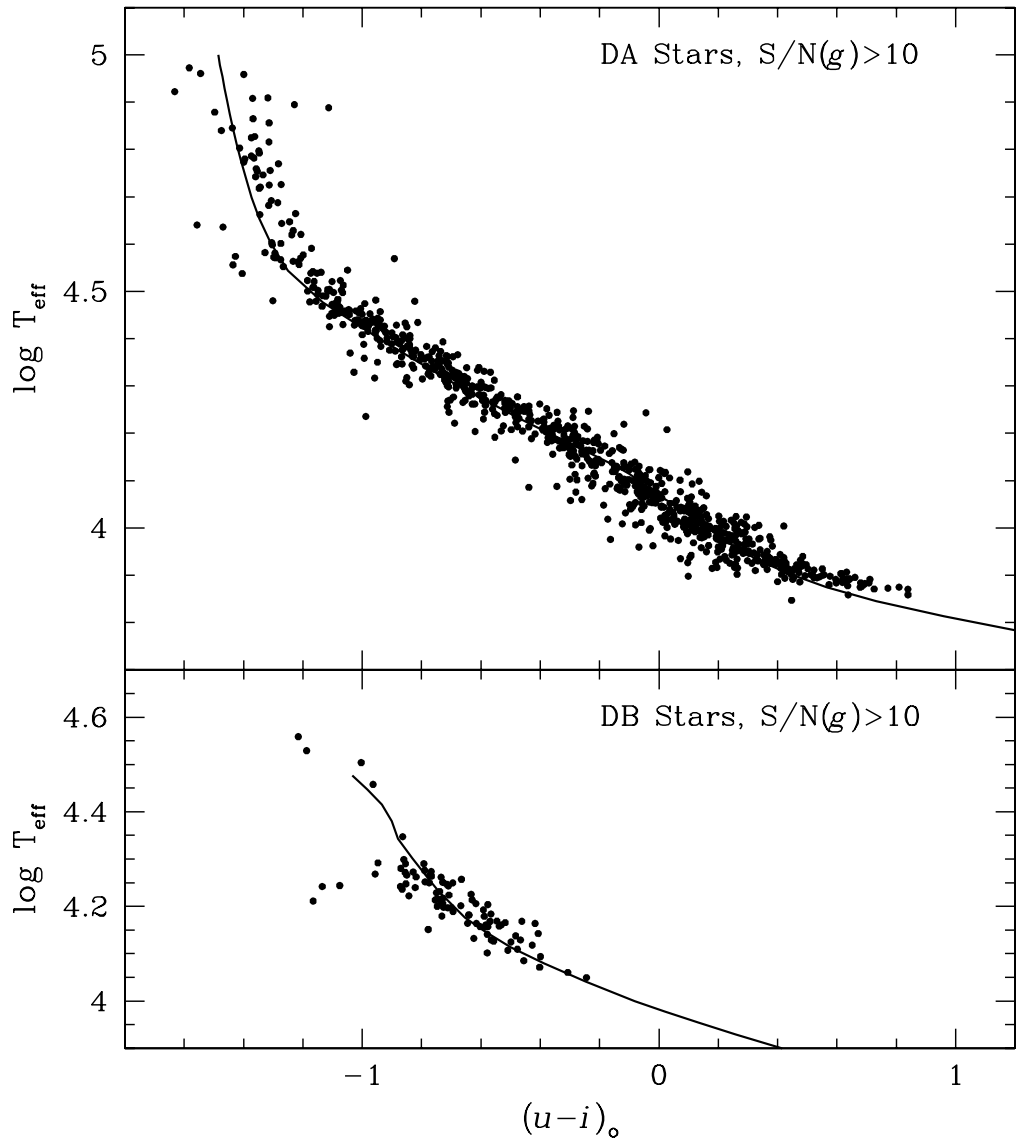}
\end{figure}

\begin{figure}
\figcaption{\label{fig:ggood}\logg vs. $(u-i)_\circ$ from our DA and DB 
model fits.  We include only data with the spectroscopic S/N in 
{\it g}$>10$/pixel.}
\plotone{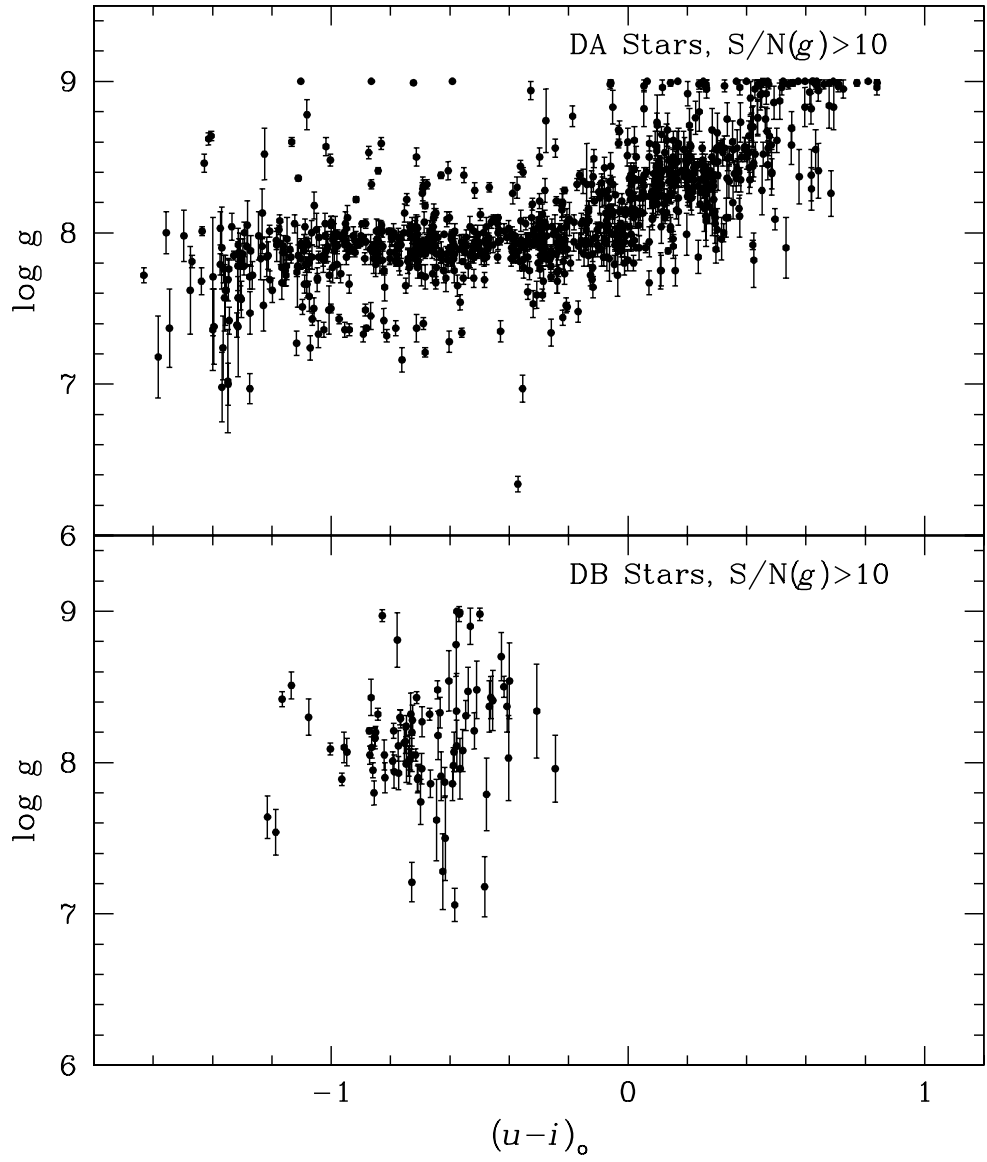}
\end{figure}

\begin{figure}
\figcaption{\label{fig:tghisto}Histograms of DA and DB \logg and \Teff fit
values.  We strongly suspect that our fits to high temperature DBs are
inaccurate and thus the few DBs with fit temperatures greater than 30,000K
are probably not so hot.}
\plotone{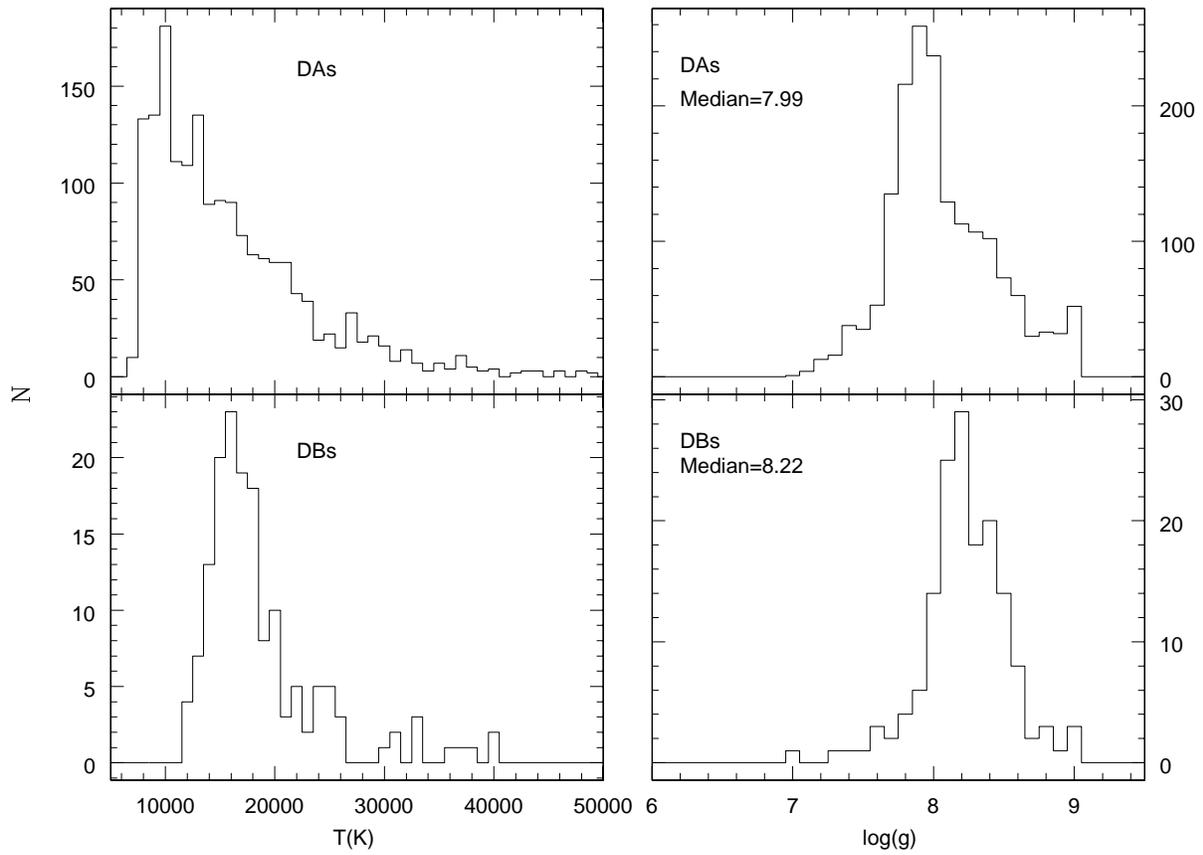}
\end{figure}

Figures~\ref{fig:tgood}~and~\ref{fig:ggood} show \Teff and \logg
vs. $(u-i)_\circ$ for our DA and DB fits.  The \Teff plot shows a nice correlation, as
expected, between \Teff and color. The line in this plot comes from
\citet{ber95} models convolved with the SDSS filters with no
AB corrections made. The AB corrections, if applied, would move the curve about
0.05 mag to the right, slightly improving the match. The fit is quite good,
although the bluest, hottest stars tend to be a little hotter than the models
predict, as discussed above.
The \logg
plot shows a mean around \loggns=8.1 (8.06 for the DAs and 8.22 for the DBs as
determined from our tables), slightly higher than results from \citet{ber92a},
for example, at \loggns=7.909.  
We also see a significant rise in \logg
for the redder objects, starting around $(u-i)_\circ=0$, corresponding to a
temperature around 12,000K.    
Figure~\ref{fig:tghisto} plots the histogram (and median) of the \logg and \Teff
fits for our DA and DB spectra  with $S/N \ge 10.0$.  This Figure includes our 
fits to all stars except those with {\it CH:} and clearly shows the bi-modal
\logg distribution seen in Figure~\ref{fig:ggood}.  The excess at \loggns=9.0 in
these plots is an artifact of our model grid which has an upper \logg limit of
9.0. 

\citet{ber90} found that the mean gravity and mass of a sample of DA
white dwarf stars cooler than the ZZ Ceti instability strip was higher than for
samples of hotter DA stars (cf. \Teff $> 15,000K$, Bergeron, Saffer \&
Liebert, 1992a).  The parameter fits for white dwarf stars cooler than
15,000K depend on the parameterization of convection (Bergeron, Wesemael,
\& Fontaine, 1992b) down to very cool temperatures where the convective
temperature gradient becomes adiabatic.  One interpretation of what
appeared to be a systematic offset to higher \logg of the cool white dwarf stars
in the Bergeron
analyses was that the mean mass was the same, but moderate amounts of
helium have been convectively mixed into the atmosphere \citep{ber90}.
Introducing helium and increasing the gravity affect
Balmer line profiles in the same way by 
increasing atmospheric pressures and densities.  Alternatively, one might
expect the coolest (and oldest) white dwarf stars to have a higher than
average mean mass since their mean progenitor mass may also
be higher.  This possibility is evaluated for the cool white dwarf sample
by Bergeron, Leggett, \& Ruiz (2001) who only show marginal evidence for
such a mass increase with decreasing temperature.
Besides, the higher mean gravity seen here seems to begin at about
11,000K where the cooling age is only about 1 Gyr.  A third possibility
for explaining the higher apparent masses and gravities of cool DA stars
is that there is a systematic error in the models --- perhaps a problem with the physics
of the hydrogen level occupation probability \citep{hum88},
or in the parameterization of convection.

\begin{figure}
\figcaption{\label{fig:typehisto}Histogram of white dwarf subclasses as a
function of $(u-i)_\circ$.}
\plotone{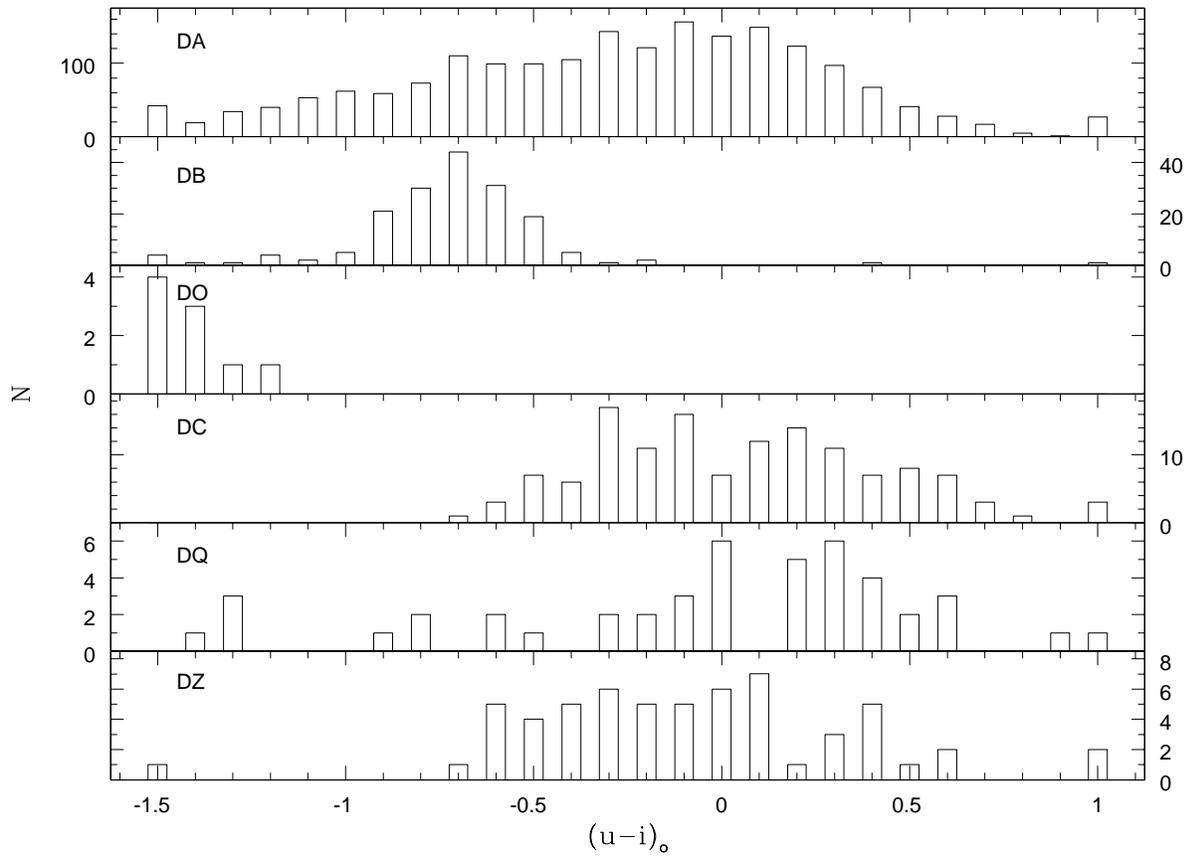}
\end{figure}

Figure~\ref{fig:typehisto} shows the object subclass classifications as a function of
$(u-i)_\circ$. The six panels show the $(u-i)_\circ$ histogram for the human classified DAs,
DBs, DOs, DCs, DQs, and DZs, respectively.  All subtypes
within each major classification and all solidly-identified spectra (regardless
of $S/N$) are included.  It is reassuring to see the DCs start at $(u-i)_\circ$
values where the DBs stop, the objects now being too cool to show He I lines.
The DOs also end where the DBs start for similar reasons.
The DQ stars, however, do overlap the DB region and are discussed further in
\citet{lie03a}. 

\begin{figure}
\figcaption{\label{fig:redpm}Reduced proper motion diagram of the white dwarf
sample with proper motions $> 12$ \masy.  The two curves are \citet{ber95} DA
models convolved with SDSS filters with assumed tangential velocities of
10 and 300 \kms.  The observed population of single white dwarf stars are
contained quite well by these two curves.}
\plotone{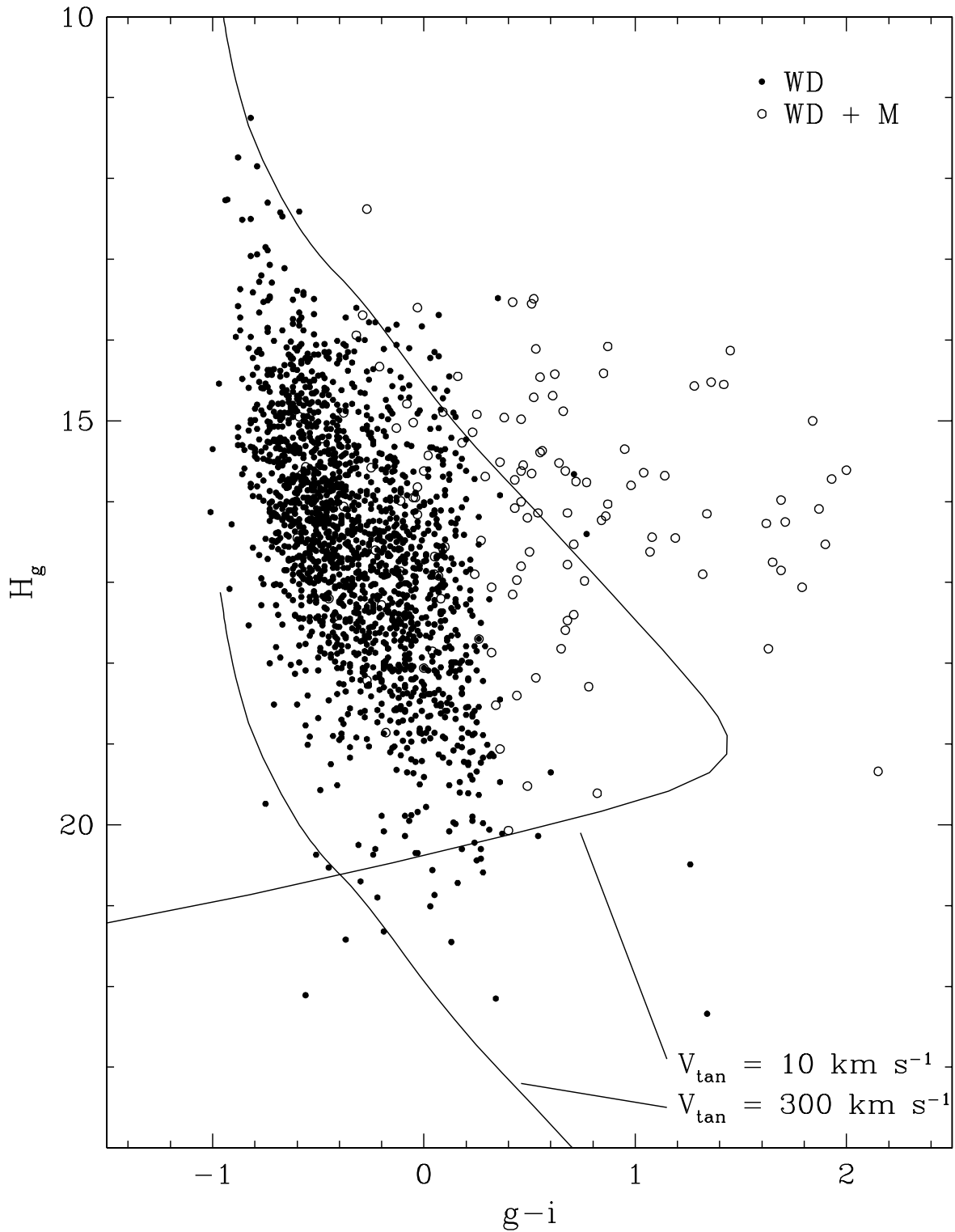}
\end{figure}

Figure~\ref{fig:redpm} is a reduced proper motion plot of our white dwarf
sample with proper motions greater than 12 \masy.  The reduced proper motion,
$H_g$, is calculated from $g+5\times
\log \mu +5$ and is meant to approximate an absolute {\it g} magnitude by
using the measured proper motion as a proxy for distance.
We plotted the white dwarf binaries (the WD+M objects) with a
different symbol since their colors will be skewed by the binary companion.
The curves show Bergeron, Wesemael, \& Beauchamp (1995) DA models (convolved
through SDSS filters) with assumed tangential velocities of 10 and
300 \kms.  As expected, the observed white dwarf population falls nicely
between these two extremes.

\section{Some Interesting Objects}

Undoubtedly in a catalog that nearly doubles the number of previously known
white dwarf stars, there are going to be some interesting, peculiar objects
contained within. \citet{har03} already discuss several such objects and we 
point out a few more here.

\subsection{Low-mass DAs}

Traditionally, astronomers have followed \citet{gre74} in defining a star with
\logg $\ge 7.0$ as a white dwarf, and lower gravity objects, subdwarf or main
sequence stars.
Existing white dwarf samples appear to include a component of low mass
stars ($<0.5$ solar), with cores composed of helium rather than
carbon-oxygen.  These may be identified from low surface gravity fits,
or from inference based on a trigonometric parallax and luminosity
determination.  Many of these candidates have been shown to be binaries,
the companion being either another white dwarf (eg. Marsh, Dhillon \& Duck
1995) or a low mass main sequence star \citep{zuc92}.

A number of stars in this catalog fit \logg $< 7.5$, which suggests a mass
near or below 0.45 solar unless the temperature is above 40,000K.  Low
mass white dwarf stars have much larger radii at high temperatures than at
lower temperatures.  Four very low mass candidates with \logg $< 7.0$
include (1) SDSS~J002207.65--101423.5 with \Teff = 19672 and \logg = 6.82, (2)
SDSS~J234536.48--010204.8 at 33049K and 6.74, (3) SDSS~J105611.03+653631.5
at 20290, 6.97, and (4) SDSS~J142601.48+010000.2 at 16465, 6.97.  Unfortunately,
the signal-to-noise ratio of the spectra allow low but less extreme
gravities in all cases, and reobservation of these stars is desirable
if the goal is to identify white dwarf stars of very low mass.

\begin{figure}
\figcaption{\label{fig:fit.335-264}Our model fit to SDSS~J123410.37-022802.9,
a probable low mass DA.}
\plotone{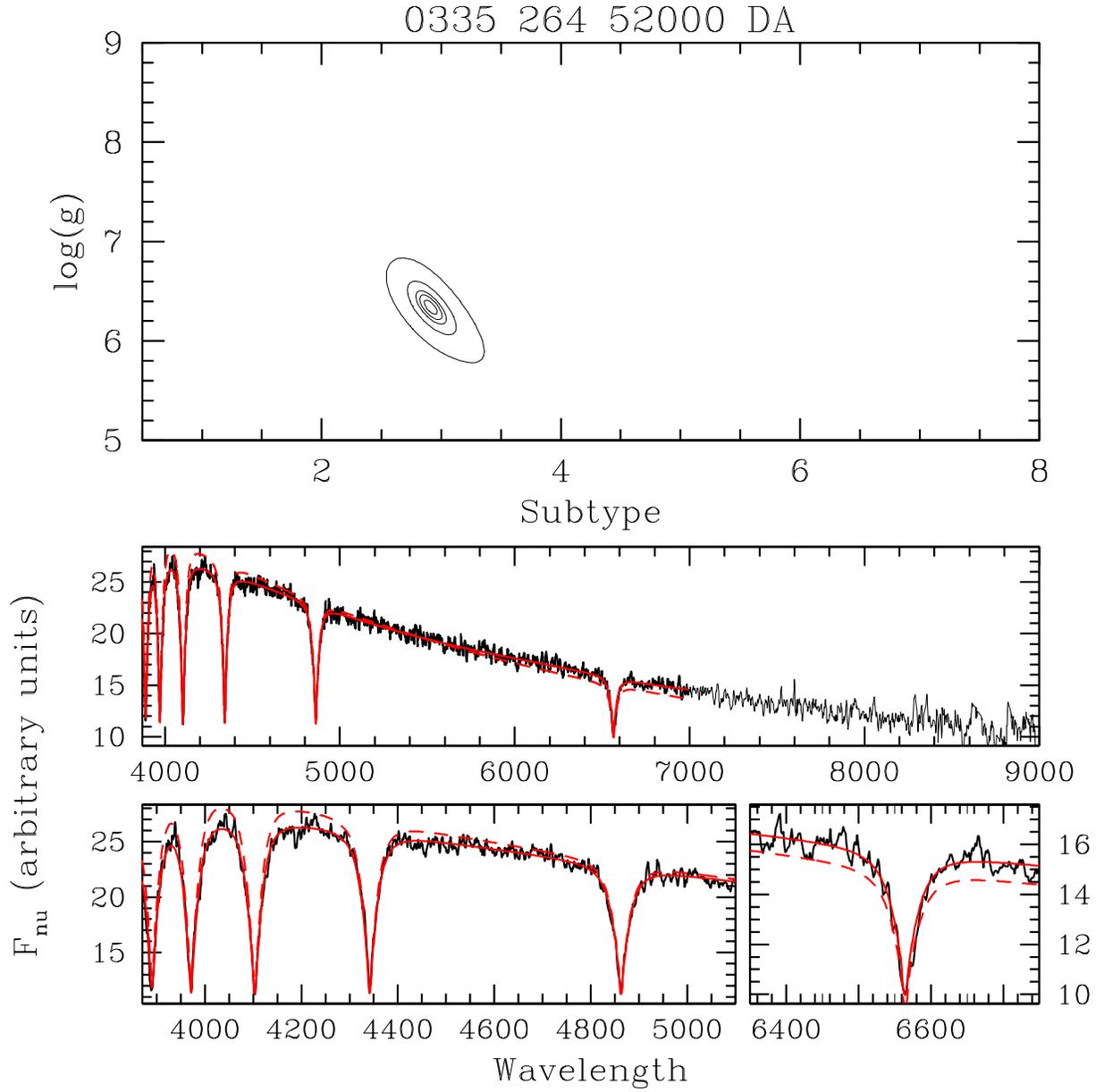}
\end{figure}

One object, SDSS~J123410.37--022802.9, has a spectrum good enough in
quality that there is no doubt about its low mass.  Our fit which measured
\Teff = 17308K and \logg = 6.34 is shown in Figure~\ref{fig:fit.335-264}.  We will defer
discussion of this so-far unique object to a separate paper \citep{lie04}.

The  DAs which we fit with \logg values 
between 6.7 and 7.0, all with significant proper motions, are listed in
Table~\ref{tb:lowmass}.

\subsection{High-mass DAs}

More than 20 DA stars shown in Figure~\ref{fig:ggood} have a fit indicating
high gravity, implying a mass significantly higher than the
average mass for white dwarf stars.  Examination of their spectra
indicates the fits are generally good and the gravity
determinations are probably correct in most cases.  Many of
the stars are faint, however, and followup spectra are
desirable.  Some of these stars have two other features that
provide support for the high gravity results:  first, their
$ugr$ colors (Figure~\ref{fig:ugr}) often lie above and to the left of the
main DA sequence as a result of their high gravities;
second, their reduced proper motions (Figure~\ref{fig:redpm}) are often large,
consistent with a low luminosity (hence closer distance) caused by a high mass
and small radius.  Occasional photometric errors or incorrect
dereddening, and the statistical nature of reduced proper
motions (strongly affected by occasional stars with high
space velocities), mean that neither of these features give a
definitive confirmation of the high gravities, only
indicative support.  Table~\ref{tb:highmass} lists our best
candidates for massive DAs with \logg$>8.5$.

\subsection{Ultra-cool White Dwarf Stars}

\citet{har01} discuss the two ultra-cool white dwarf stars found
by the SDSS. One, SDSS~J133739.40+000142.9, is a unique discovery of the SDSS
while the second, SDSS~J165401.26+625355.0, was previously known as LHS~3250 and
is discussed in \citet{har99}.  Both are believed to have low masses
and helium cores.  LHS~3250 has a
measured parallax, and \citet{ber02} find that including
the parallax gives a solution with \logg = 7.27 and a mass of 0.23 \Msun.
We found no more similar objects in the expanded DR1 data set.

It thus appears that the LHS~3250-like white dwarf stars, with very strong
opacity due to collisionally-induced absorption (CIA) bands of molecular
hydrogen, are quite rare.  \citet{ber02} showed
quantitatively that LHS~3250 is a very low mass helium object.  In the
discovery paper, \citet{har01} pointed out that it is
kinematically a likely member of the disk, not the halo population.
Indeed, \citet{alt01} and \citet{ser01} demonstrate
that low mass helium core white dwarf stars (with hydrogen envelopes) 
evolve into the regime where the CIA opacity becomes dominant in the
infrared to I band in a shorter time than do white dwarf stars of normal
mass (ie. arguably, within the age of the disk, not the halo).  The SDSS
sample has uncovered no candidates of more normal mass from the Galactic
halo.

One caveat has to be added, however.  The very cool candidate
WD~0346+246 discussed by \citet{hod00} {\it does} have halo kinematics,
but it may not show the CIA opacity strongly enough to be pulled out of
the stellar locus were it observed by the SDSS --- which is how LHS~3250 and
SDSS~1337 were found.
Hence, the conclusion of the previous paragraph does not rule out the
presence of many cool, halo white dwarf stars in the SDSS imaging data.

\subsection{Hot DBs?}

There are 12 DBs in Table~\ref{tb:DBs} that fit to a \Teff $>30,000K$.
Since no DB has yet to be found above 30,000K, the so-called ``DB gap'' 
\citep{lie86}, these are very interesting objects!

Two of these objects, SDSS~J163838.24--005417.5 and SDSS~J040854.60--043354.7
show obvious signs of hydrogen in their spectra while~SDSS J090456.13+525029.9
shows perhaps a weak hint of H$\alpha$.  The fit to SDSS~J102615.54+005942.9
has two minima and the spectrum is quite noisy (as indicated by the large
uncertainty in the \Teff determination), so it may just be that the proper
fit is the cooler, second minimum.
In addition, SDSS~J203517.50-064539.9's fit includes a {\it :} which
arose from a bad match between the fit spectrum and the SDSS photometry.

The remaining seven objects, however, seem to have acceptable fits, no
visible sign of hydrogen in their spectra and colors which generally agree
with a \Teff at or above 30,000K. Given the systematic uncertainties
in our model fitting, as well as the noise in these spectra, we do not claim
these objects as certain occupants of the DB gap. Better spectra and more
careful fitting will be required to assess
these remaining, potentially very intriguing hot DBs.

\subsection{An Eclipsing, Pre-Cataclysmic Variable?}
 
The DA SDSS~J010622.99--001456.3 is another interesting case. Its {\it best}
photometric magnitudes as listed in our tables indicate a very dim red 
object, ({\it u,g,r,i,z})=(25.3, 24.8, 20.7, 20.8, 19.5),
whereas the spectrum appears like a substantially brighter 11,000K DA.
These $u$ and $g$ magnitudes are basically non-detections.
The SDSS {\it best} image of the object looks fine except for the presence
of a very dim, red object instead of the brighter, bluer one we would expect to
see.  The {\it target} photometry, ({\it u,g,r,i,z})=(18.6, 18.1, 18.3, 18.5, 18.3), shows it to be
a normal, bluish, bright DA, in good agreement with the spectrum. 
We actually have six separate imaging observations of this object and in all but
one case (the one ironically labeled {\it best}), the magnitudes agree with
both the {\it target} data and the spectrum.

The spectrum does show some possible
(although noisy) excess in the red, perhaps indicating a faint red
companion, which if it occasionally produces a total eclipse of the
white dwarf star, could explain the one deviant set of photometric
measurements.  Our fit to the spectrum supports this idea; the white
dwarf appears bluer than the model.  The {\it best} $(r-i)$ and $(i-z)$
magnitudes are not quite right for a typical late-type star, but there is a
time delay between each filter measurement, so we could be seeing some of the
white dwarf in one filter and more or less in another, as the eclipse
progresses.

The previously-known object PG1413+015 (GK~Vir: Green et al. 1978, 
Fulbright et al. 1993, Green, Schmidt, \& Liebert 1986) is an eclipsing binary
system consisting of a DAO white dwarf star with a roughly M4~V companion.  
Its orbital period is $8^{\rm h} 16^{\rm m}$ which places it among the shortest period white
dwarf binaries that do not show signs of interaction with its companion.
The natural evolution of this system is thought to lead to a cataclysmic
variable and thus it represents a rare pre-cataclysmic
system.  Whether or not SDSS~J010622.99--001456.3 is a similar system
will require future observations, but current observations are consistent
with such an interpretation. We are now planning to observe this star
periodically for radial velocity and color changes.

\subsection{One Non-Cataclysmic-Variable}

The star SDSS~J131751.72+673159.4 is classified here as DAME,
showing sharp H and HeI emission lines in the cores of the
broader absorption lines of the white dwarf.  The emission lines
are sharper than those seen in the numerous SDSS spectra of
cataclysmic variables (CVs) \citep{szk02,szk03}.  This star must have a close
cool companion, and may become a CV, but it is not a CV now.
It is listed in the online version of the \citet{dow01} CV Catalog,
where we suggest it should be deleted.

\subsection{DAs with Weak Balmer Lines}

Three interesting stars are classified as DA: in Table~\ref{tb:unc}:
SDSS~J102448.85-002312.3, SDSS J150856.89+013557.8, and
SDSS~J164306.06+442638.1.  They all have significant proper motions,
colors indicating temperatures between about 10000--14000 K, and weak
H$\alpha$ and possibly higher Balmer lines.  If they were
normal DA white dwarf stars, they would have much stronger Balmer
lines.  They are likely to be white dwarf stars with atmospheres
dominated by helium but including a small amount of hydrogen.
They would appear as DBA stars if they were a little hotter,
or DC stars if they were cooler.

\section{Conclusions}

By selecting candidate spectra based mostly on a variety of color and proper motion
cuts, we have found 2791 spectroscopically-confirmed white dwarf and subdwarf
stars in the first data release of the SDSS.  We currently see no reason why
this detection rate should change in future SDSS data releases,  and thus can
look forward to an additional 10,000 or so white dwarf stars from the SDSS
by the time it is finished.  

This and future catalogs will allow us to expand dramatically our sample
of particularly unique, interesting, and rate white dwarf stars. We have
already pointed out some of the work going on in this regard with the DQ, DH,
pulsating, and hot DO white dwarf stars.  We expect to make updated catalogs
available as more data are released by the SDSS and welcome contact with
fellow white dwarf researchers with regards to SDSS objects.

An ultimate goal of white dwarf research is a vastly improved luminosity
function over what is known at present. The vagaries of how SDSS white dwarf
spectra are obtained makes building a full luminosity function a difficult task,
but one which should be ultimately possible.  However, a luminosity function
concentrating on the hot end of the white dwarf cooling sequence suffers from
fewer targeting complications and thus should be doable on much shorter
timescales than the entire luminosity function will require. We are currently
beginning this work in addition to preparing the next public catalog of white
dwarf stars from the SDSS.  The data tables presented here are available online
at the SDSS DR1 value added catalog site: 
{\url http:///www.sdss.org/dr1/products/value\_added/wdcat/dr1/index.html}.

\acknowledgments

We are grateful to Detlev Koester for providing us his fine grid of DA and DB
models we used in our model fits.  SJK, AN, and JK would like to thank
J. Peoples and B. Gillespie for their support of our research effort.
D.J.E. was supported by NSF AST-0098577 and an Alfred P. Sloan Research 
Fellowship.

Funding for the creation and distribution of the SDSS Archive has been provided
by the Alfred P. Sloan Foundation, the Participating Institutions, the National
Aeronautics and Space Administration, the National Science Foundation, the U.S.
Department of Energy, the Japanese Monbukagakusho, and the Max Planck Society.
The SDSS Web site is \url{http://www.sdss.org/}.

The SDSS is managed by the Astrophysical Research Consortium (ARC) for the
Participating Institutions. The Participating Institutions are The University
of Chicago, Fermilab, the Institute for Advanced Study, the Japan Participation
Group, The Johns Hopkins University, Los Alamos National Laboratory, the
Max-Planck-Institute for Astronomy (MPIA), the Max-Planck-Institute for
Astrophysics (MPA), New Mexico State University, University of Pittsburgh,
Princeton University, the United States Naval Observatory, and the University
of Washington.

\clearpage

\appendix
\section{Comparison with the McCook and Sion White Dwarf Catalog}

Matching the stars in this paper with those in the online version of the
McCook \& Sion (1999) catalog combined with a preliminary list of WDs found in
the Hamburg Quasar Survey \citep{hom98} finds 109 white dwarf stars (one of
which we find to be a sdB star) in this paper
have been already published, not counting those SDSS WDs already reported by
\citet{har03} and \citet{ray03}.  These previously known stars are listed in
Table~\ref{tb:known}.  Not surprisingly, these stars tend to be relatively
bright and our spectral classifications are generally consistent with those
published previously.  Different SDSS positions compared to the previously
published positions
noted for some stars in the table can usually be understood by the stars'
proper motions and/or by truncation or imprecise original coordinates ---
the SDSS coordinates should be used.

We also searched the McCook and Sion (1999) catalog  for white dwarf stars that
fell within the 2099 deg$^2$ of the DR1 imaging sky area, but which were not recovered by us.  Remember,
the SDSS targeting of white dwarf stars for spectroscopy is rather haphazard;
most of our spectra come from targeting categories other than those searching
for white dwarf stars and are thus ``rejects'' of these other categories.
However, we find only 218 known white dwarf stars within the DR1 imaging area for
which we
do not have spectra. This number includes objects in the 769 deg$^2$ of DR1
imaging area that do not currently have DR1 spectra.  Some of these objects
may still get SDSS spectra that will become available in a future data release.
uture objects, along with the SDSS PSF magnitudes and
errors, are shown in Table~\ref{tb:notindr1}. The coordinates listed are those
from the SDSS.  Since many known white dwarf stars are too bright for the SDSS
to measure accurately, any magnitude which is flagged as {\it SATURATED} in the
DR1 database is marked with an {\it *} next to the listed magnitude. SDSS
saturated magnitudes are not reliable.


\clearpage

\clearpage

\clearpage

\clearpage

\clearpage

\clearpage

\clearpage

\clearpage

\clearpage

\clearpage

\clearpage

\clearpage
 
\clearpage

\clearpage


\begin{deluxetable}{lrrrrrrrrrr}
\tablecaption{\label{tb:objtypes}SDSS targeting categories for all
presented white dwarf and hot subdwarf spectra.}


\tabletypesize{\small}
\tablewidth{0pt}
\tablehead {
\colhead{OBJTYPE} & \colhead{DAs} & \colhead{DBs} & \colhead{DOs} &
\colhead{DCs} & \colhead{DQs} & \colhead{DZs} & \colhead{DHs} &
\colhead{WDM} & \colhead{SDs} & \colhead{ALL} }
\startdata
SERENDIPITY\_MANUAL       &   3 &  0 &   0 &   0 &   0 &   0 &   0 &   1 &   0 &    4 \\
GALAXY                    &   0 &  0 &   0 &   0 &   0 &   0 &   0 &   3 &   2 &    5 \\
ROSAT                     &   9 &  1 &   0 &   0 &   0 &   0 &   0 &   0 &   0 &   10 \\
QA                        &   8 &  0 &   0 &   1 &   0 &   0 &   0 &   2 &   0 &   11 \\
STAR\_BHB                 &  19 &  0 &   0 &   0 &   0 &   1 &   0 &   0 &   5 &   25 \\
STAR\_CATY\_VAR           &  15 &  1 &   0 &   2 &   0 &   0 &   0 &  17 &   0 &   35 \\
STAR\_WHITE\_DWARF        & 131 &  8 &   0 &   3 &   1 &   2 &   4 &   1 &   7 &  157 \\
SERENDIPITY\_DISTANT      & 417 & 34 &   0 &   6 &   2 &   6 &   4 &   1 &  34 &  505 \\
HOT\_STD                  & 268 & 85 &  13 &   7 &   7 &   9 &   7 &  11 & 124 &  531 \\
QSO                       & 390 &  3 &   0 &  62 &  31 &  21 &  10 & 133 &  50 &  707 \\
SERENDIPITY\_BLUE         & 628 & 39 &   0 &  53 &   4 &  18 &   5 &  31 &  18 &  801 \\
\enddata

\tablecomments{The numbers in the the individual category
columns do not sum to those in the ALL
column since the latter includes the certain white dwarf stars,
but of uncertain subclass.}

\end{deluxetable}

\begin {deluxetable}{lllr}
\tablecaption{\label{tb:selection}Number of resultant objects, N, and criteria for
candidate spectra selection.}


\tablewidth{0pt}
\tablehead {
\colhead{Classification} & \colhead{Photometric} & \colhead{Other} & 
\colhead{N} \\
\colhead{Type} & \colhead{Constraints} & \colhead{Constraints} }
\startdata
Blue & $(u-g)_\circ < 0.70$ & specClass != 3 & 4937 \\*
     & $(g-r)_\circ < -0.10$ & (ie. object not a QSO) & \\*
     & $u < 21.0$ &  & \\[1ex]

Medium Blue & $(u-g)_\circ < 0.60$ & specClass = 0,1 or 6  & 3553 \\*
            & $(g-r)_\circ > -0.10$ &  (star or unknown) OR & \\*
            & $u < 21.0$ & Zstatus = 0 or 1 or 2 & \\*
            &            & (z measurement failed or inconsistent) OR & \\*
            &            & z $< 0.01$ & \\[1ex]

Medium Red & $(u-g)_\circ > 0.60$ & objc\_type\tablenotemark{1} = STAR & 537 \\*
w/ Proper Motion & $(g-r)_\circ < 1.00$ & $0.8'' < \delta  < 10.0''$ 
                                        ~\tablenotemark{2} \\*
          & $H_g > 17.0$ ~\tablenotemark{3} ~~OR & & \\*
          & $H_g> 14.0 + 3.0*(g-i)_\circ$ && \\[1ex]

Targeted WDs & & STAR\_WHITE\_DWARF target flag set & 1575 \\[1ex]

Princeton\tablenotemark{4}~~WD & & spectrum classified as WD, DH, DQ OR & 4122 \\*
or HOT Star &  & as O, OB, B6, B9, A0, A0p by alternate \\*
            &  &  pipeline at Princeton& \\*
            & $(u-g)_\circ<0.90$ if ---$>$& classified as B or A & \\*
            & (to exclude BHB) & \\[1ex]

Eyeball & & rejects from other searches & 1138 \\*
        & & with manual possible white dwarf ID made & \\

\enddata
\tablenotetext{1}{objc\_type is a {\it frames} pipeline output that does a simple star/galaxy
separation.  See \url{http://www.sdss.org/dr1/algorithms/classify.html}.}
\tablenotetext{2}{$\delta$ is the difference between USNO-A catalog coordinates
and SDSS observed coordinates, available in the USNO table in the SDSS
databases. The timebase is about 50 years.}
\tablenotetext{3}{$H_g$ is the reduced proper motion: $H_g = g + 5 \times
\log \mu +5$ where $\mu$ is the proper motion in $"/yr$.}
\tablenotetext{4}{D. Schlegel (private communication) has an independent
spectroscopic pipeline, specBS, running at Princeton. This selection category
uses results from that pipeline.}

\end{deluxetable}

\begin {deluxetable}{lllrrrrrrrrrrrrrll}
\tablecaption{\label{tb:DAs}SDSS DR1 spectroscopically identified DA stars.}
\tablecomments{Table~\ref{tb:DAs} is published in its entirety in the
electronic edition of the {\it Astrophysical Journal} and at the URL provided
in the text. A portion is shown here
for guidance regarding its form and content.  The SDSS Object name, and the
{\it g,r,i,z} magnitudes have been removed from this sample table which
displays the data from the first and last three entries of the full table.}
\rotate
\tabletypesize{\scriptsize}
\tablewidth{0pt}
\tablehead{
\colhead{Plate} & \colhead{MJD} & \colhead{Fiber} &
\colhead{RA} & \colhead{Dec} &
\colhead{u$_{psf}$} & \colhead{$\delta_u$} &
\colhead{S/N$_g$} &
\colhead{PM$_{RA}$} & \colhead{PM$_{Dec}$} &
\colhead{A$_g$} &
\colhead{$\rm T_{eff}$} & \colhead{$\delta_{\rm T_{eff}}$} &
\colhead{$\rm \log g$} & \colhead{$\delta_{\rm \log g}$} &
\colhead{$chi^2$} & \colhead{AutoID} &
\colhead{Notes}\\
\colhead{} &
\colhead{} &
\colhead{} &
\colhead{(deg)} & \colhead{(deg)} &
\colhead{} &
\colhead{} &
\colhead{} &
\colhead{(mas~yr$^{-1}$)} & \colhead{(mas~yr$^{-1}$)} &
\colhead{} &
\colhead{(K)} & \colhead{(K)}
}
\startdata
                               
  0651 & 52141 & 382 & 1.34173 & --9.54185 & 20.30 & 0.06 &                 3.61 & 18 & --10 & 0.142 & 0 & 0 & 0.00 & 0.00 & 0.00 & N/A & 2\\
  0388 & 51793 & 001 & 2.67918 & --0.89987 & 19.41 & 0.03 &                 5.78 & --26 & --12 & 0.154 & 0 & 0 & 0.00 & 0.00 & 0.00 & N/A & 2\\
  0389 & 51795 & 431 & 3.41294 & 0.32358 & 15.76 & 0.02 &                 70.39 & 398 & --175 & 0.107 & 0 & 0 & 0.00 & 0.00 & 0.00 & N/A & 2\\
  0366 & 52017 & 151 & 262.45991 & 58.30243 & 18.29 & 0.02 &                 11.32 & 4 & --8 & 0.159 & 90975 & 6897 & 7.71 & 0.39 & 1.01 & DA0.6\_7.7\\
  0494 & 51915 & 052 & 188.65494 & 66.72683 & 18.40 & 0.02 &                 15.09 & --1 & --4 & 0.056 & 91304 & 6285 & 7.37 & 0.26 & 0.97 & DA0.6\_7.4\\
  0413 & 51929 & 483 & 49.74284 & 0.39049 & 17.91 & 0.02 &                 16.89 & 3 & --2 & 0.300 & 93855 & 5768 & 7.18 & 0.27 & 1.00 & DA0.5\_7.2\\
\enddata

\tablenotetext{1}
 {Computer fit checked and found to be a reasonable fit.}
\tablenotetext{2}
 {Computer fit checked and found not to be a believable fit.}
\tablenotetext{3}
 {The computer fit may be affected by a late-type companion.}
\tablenotetext{4}
 {This particular object is described further in the ``Interesting
 Objects'' section of the text.}
\tablenotetext{5}
 {This object had no ``best'' photometry in the DR1 database, so the
 ugriz magnitudes come from the ``target'' photometry.  (The lack
 of ``best'' photometry is probably due to the object being on the
 edge of the DR1 survey limits.)}
\tablenotetext{6}
 {This DB spectrum also shows signs of Hydrogen.}
\tablenotetext{*}
 {Magnitudes marked with an * have bad photometric pipeline quality
 control flags set.}

\end{deluxetable}

\begin {deluxetable}{lllrrrrrrrrrrrrrll}
\tablecaption{\label{tb:DBs}SDSS DR1 spectroscopically identified DB stars.}
\tablecomments{Table~\ref{tb:DBs} is published in its entirety in the
electronic edition of the {\it Astrophysical Journal} and at the URL provided
in the text. A portion is shown here
for guidance regarding its form and content.  The SDSS Object name, and the
{\it g,r,i,z} magnitudes have been removed from this sample table which
displays the data from the first and last three entries of the full table.}
\rotate
\tabletypesize{\scriptsize}
\tablewidth{0pt}
\tablehead{
\colhead{Plate} & \colhead{MJD} & \colhead{Fiber} &
\colhead{RA} & \colhead{Dec} &
\colhead{u$_{psf}$} & \colhead{$\delta_u$} &
\colhead{S/N$_g$} &
\colhead{PM$_{RA}$} & \colhead{PM$_{Dec}$} &
\colhead{A$_g$} &
\colhead{$\rm T_{eff}$} & \colhead{$\delta_{\rm T_{eff}}$} &
\colhead{$\rm \log g$} & \colhead{$\delta_{\rm \log g}$} &
\colhead{$chi^2$} & \colhead{AutoID} &
\colhead{Notes}\\
\colhead{} &
\colhead{} &
\colhead{} &
\colhead{(deg)} & \colhead{(deg)} &
\colhead{} &
\colhead{} &
\colhead{} &
\colhead{(mas~yr$^{-1}$)} & \colhead{(mas~yr$^{-1}$)} &
\colhead{} &
\colhead{(K)} & \colhead{(K)}
}
\startdata
                               
  0427 & 51900 & 320 & 29.12326 & 13.29576 & *17.95 & 0.02 &                 23.04 & 8 & --4 & 0.211 & 0 & 0 & 0.00 & 0.00 & 0.00 & N/A & 2\\
  0455 & 51909 & 247 & 38.38922 & --9.39038 & 18.10 & 0.03 &                 18.17 & 0 & 4 & 0.107 & 0 & 0 & 0.00 & 0.00 & 0.00 & N/A & 2\\
  0464 & 51908 & 278 & 59.07291 & --6.51934 & 18.68 & 0.02 &                 16.23 & 13 & 17 & 0.287 & 0 & 0 & 0.00 & 0.00 & 0.00 & N/A & 2\\
  0532 & 51993 & 358 & 210.49622 & 2.35743 & 18.73 & 0.03 &                 10.41 & --2 & 8 & 0.120 & 38211 & 1188 & 7.92 & 0.17 & 1.17 & DB1.3\_7.9\\
  0465 & 51910 & 518 & 62.22750 & --4.56518 & 19.05 & 0.03 &                 10.64 & --3 & --17 & 0.364 & 40000 & 1016 & 8.09 & 0.17 & 1.23 & DB1.3\_8.1: & 4~6\\
  0348 & 51671 & 003 & 249.65933 & --0.90486 & 19.42 & 0.03 &                 8.33 & --1 & --16 & 0.506 & 40000 & 317 & 7.05 & 0.06 & 1.45 & DB1.3\_7.1: & 4~6\\
\enddata

\tablenotetext{1}
 {Computer fit checked and found to be a reasonable fit.}
\tablenotetext{2}
 {Computer fit checked and found not to be a believable fit.}
\tablenotetext{3}
 {The computer fit may be affected by a late-type companion.}
\tablenotetext{4}
 {This particular object is described further in the ``Interesting
 Objects'' section of the text.}
\tablenotetext{5}
 {This object had no ``best'' photometry in the DR1 database, so the
 ugriz magnitudes come from the ``target'' photometry.  (The lack
 of ``best'' photometry is probably due to the object being on the
 edge of the DR1 survey limits.)}
\tablenotetext{6}
 {This DB spectrum also shows signs of Hydrogen.}
\tablenotetext{*}
 {Magnitudes marked with an * have bad photometric pipeline quality
 control flags set.}

\end{deluxetable}

\begin {deluxetable}{lllrrrrrrrrrll}
\tablecaption{\label{tb:all}All SDSS DR1 spectroscopically identified white
dwarf and subdwarf stars.}
\tablecomments{Table~\ref{tb:all} is published in its entirety in the
electronic edition of the {\it Astrophysical Journal} and at the URL provided
in the text. A portion is shown here
for guidance regarding its form and content.  The SDSS Object name, and the
{\it g,r,i,z} magnitudes have been removed from this sample table which
displays the data from the first and last three entries of the full table.}
\rotate
\tabletypesize{\scriptsize}
\tablewidth{0pt}
\tablehead{
\colhead{Plate} & \colhead{MJD} & \colhead{Fiber} &
\colhead{RA} & \colhead{Dec} &
\colhead{Epoch} &
\colhead{u$_{psf}$} & \colhead{$\delta_u$} &
\colhead{S/N$_g$} &
\colhead{PM$_{RA}$} & \colhead{PM$_{Dec}$} &
\colhead{A$_g$} &
\colhead{Human ID} &
\colhead{Notes}\\
\colhead{} &
\colhead{} &
\colhead{} &
\colhead{(deg)} & \colhead{(deg)} &
\colhead{} &
\colhead{} &
\colhead{} &
\colhead{} &
\colhead{(mas~yr$^{-1}$)} & \colhead{(mas~yr$^{-1}$)} &
}
\startdata
                                 
  0650 & 52143 & 497 & 0.02983 & --9.72773 & 2000.74 & 19.40 & 0.04 &                 7.74 & N/A & N/A & 0.127 & DA\\
  0650 & 52143 & 450 & 0.04820 & --8.83566 & 2000.74 & 19.42 & 0.04 &                 11.04 & 106 & --2 & 0.140 & DQ\\
  0650 & 52143 & 217 & 0.09390 & --10.86172 & 2000.74 & 19.28 & 0.03 &                 12.17 & 43 & --27 & 0.131 & DA5\\
  0386 & 51788 & 035 & 358.87400 & --0.00391 & 2001.79 & 20.14 &                 0.12 & 4.67 & 1 & --2 & 0.150 & DA\\
  0387 & 51791 & 347 & 359.22201 & 0.36072 & 2001.79 & 19.72 &                 0.05 & 8.76 & 76 & 85 & 0.141 & DA\\
  0650 & 52143 & 233 & 359.60751 & --10.57039 & 2000.74 & 17.25 &                 0.04 & 35.13 & 59 & --21 & 0.122 & DA\\
\enddata

\tablenotetext{1}
 {Computer fit checked and found to be a reasonable fit.}
\tablenotetext{2}
 {Computer fit checked and found not to be a believable fit.}
\tablenotetext{3}
 {The computer fit may be affected by a late-type companion.}
\tablenotetext{4}
 {This particular object is described further in the ``Interesting
 Objects'' section of the text.}
\tablenotetext{5}
 {This object had no ``best'' photometry in the DR1 database, so the
 ugriz magnitudes come from the ``target'' photometry.  (The lack
 of ``best'' photometry is probably due to the object being on the
 edge of the DR1 survey limits.)}
\tablenotetext{6}
 {This DB spectrum also shows signs of Hydrogen.}
\tablenotetext{*}
 {Magnitudes marked with an * have bad photometric pipeline quality
 control flags set.}

\end{deluxetable}

\begin {deluxetable}{lllrrrrrrrrrll}
\tablecaption{\label{tb:unc}All SDSS DR1 spectroscopically identified white 
dwarf and subdwarf stars with uncertain identifications.}
\tablecomments{Table~\ref{tb:unc} is published in its entirety in the
electronic edition of the {\it Astrophysical Journal} and at the URL provided
in the text. A portion is shown here
for guidance regarding its form and content.  The SDSS Object name, and the
{\it g,r,i,z} magnitudes have been removed from this sample table which
displays the data from the first and last three entries of the full table.}
\rotate
\tabletypesize{\scriptsize}
\tablewidth{0pt}
\tablehead{
\colhead{Plate} & \colhead{MJD} & \colhead{Fiber} &
\colhead{RA} & \colhead{Dec} &
\colhead{Epoch} &
\colhead{u$_{psf}$} & \colhead{$\delta_u$} &
\colhead{S/N$_g$} &
\colhead{PM$_{RA}$} & \colhead{PM$_{Dec}$} &
\colhead{A$_g$} &
\colhead{Human ID} &
\colhead{Notes}\\
\colhead{} &
\colhead{} &
\colhead{} &
\colhead{(deg)} & \colhead{(deg)} &
\colhead{} &
\colhead{} &
\colhead{} &
\colhead{} &
\colhead{(mas~yr$^{-1}$)} & \colhead{(mas~yr$^{-1}$)} &
}
\startdata
                                 
  0388 & 51793 & 350 & 1.29140 & 0.63600 & 2001.79 & 17.77 & 0.03 &                 16.37 & 23 & 23 & 0.157 & SDO:\\
  0417 & 51821 & 359 & 7.05945 & 15.02500 & 1999.78 & 20.84 & 0.08 &                 2.94 & N/A & N/A & 0.246 & DA:\\
  0417 & 51821 & 345 & 7.13435 & 15.23070 & 1999.78 & 20.59 & 0.07 &                 4.45 & N/A & N/A & 0.254 & SDB:\\
  0386 & 51788 & 403 & 357.27549 & 1.00614 & 2002.68 & 19.94 &                 0.20 & 3.73 & 0 & --16 & 0.097 & DB:\\
  0386 & 51788 & 054 & 358.12767 & --0.64107 & 2002.68 & 20.34 &                 0.48 & 2.42 & --5 & --11 & 0.111 & DA:\\
  0386 & 51788 & 110 & 358.59147 & --0.35811 & 2002.68 & 20.71 &                 0.19 & 3.77 & 2 & --8 & 0.116 & DA:\\
\enddata

\tablenotetext{1}
 {Computer fit checked and found to be a reasonable fit.}
\tablenotetext{2}
 {Computer fit checked and found not to be a believable fit.}
\tablenotetext{3}
 {The computer fit may be affected by a late-type companion.}
\tablenotetext{4}
 {This particular object is described further in the ``Interesting
 Objects'' section of the text.}
\tablenotetext{5}
 {This object had no ``best'' photometry in the DR1 database, so the
 ugriz magnitudes come from the ``target'' photometry.  (The lack
 of ``best'' photometry is probably due to the object being on the
 edge of the DR1 survey limits.)}
\tablenotetext{6}
 {This DB spectrum also shows signs of Hydrogen.}
\tablenotetext{*}
 {Magnitudes marked with an * have bad photometric pipeline quality
 control flags set.}

\end{deluxetable}

\begin{deluxetable}{rl}
\tablecaption{\label{tb:badflags}SDSS photometric pipeline output flags used to
indicate bad photometry.}
\tablewidth{0pt}
\tablehead{ \colhead{Flag} & \colhead{Description}}
\startdata
EDGE &  object too close to edge of frame to be measured \\
PEAKCENTER & used brightest pixel as centroid \\
NOPROFILE & only 0 or 1 entries for the radial flux profile \\
BAD\_COUNTS\_ERROR & interpolation affected many pixels \\
INTERP\_CENTER  & interpolated pixel(s) within 3 pixels of the center \\
& (we only use this flag if a cosmic ray was flagged as well)\\
DEBLEND\_NOPEAK & object is a CHILD of a DEBLEND but has no peak \\
& (we only use this flag if the PSF magnitude error $>$ 0.2 mag) \\
PSF\_FLUX\_INTERP & more than 20\% of PSF flux is interpolated over \\
SATUR & contains saturated pixels \\
NOTCHECKED & object contains pixels which were not checked for peaks by deblender \\
\enddata
\end{deluxetable}

\begin{deluxetable}{lr}
\tablecaption{\label{tb:reddening}Multiplicative conversion values, N, to go 
from $A_g$ to the extinction in any other SDSS filter.  $A_{\it x} = N*A_g$.}
\tablewidth{0pt}
\tablehead{ \colhead{Filter} & \colhead{N}}
\startdata
{\it u} & 1.360 \\
{\it g} & 1.000 \\
{\it r} & 0.726 \\
{\it i} & 0.550 \\
{\it z} & 0.390 \\
\enddata
\end{deluxetable}

\clearpage

\begin{deluxetable}{lrr}
\tablecaption{\label{tb:lowmass}Fitted DAs with measured \logg values
between 6.7 to 7.0.}
\tablewidth{0pt}
\tablehead{ \colhead{Name} & \colhead{\logg} &\colhead{\Teff}}
\startdata
SDSS~J123410.37--022802.9 & $6.34 \pm 0.05$ & $17308 \pm 226$ \\
SDSS~J234536.48--010204.8 & $6.74 \pm 0.23$ & $33049 \pm 1105$ \\
SDSS~J002207.65--101423.5 & $6.82 \pm 0.16$ & $19672 \pm 729$ \\
SDSS~J105611.03+653631.5 &  $6.97 \pm 0.12$ & $20290 \pm 637$ \\
SDSS~J131033.26+644032.9 & $6.97 \pm 0.10$ & $39937 \pm 918$ \\
SDSS~J142601.48+010000.2 &  $6.97 \pm 0.09$ & $16465 \pm 355$\\
SDSS~J163800.36+004717.8 & $6.98 \pm 0.23$ & $73256 \pm 4366$ \\
\enddata
\end{deluxetable}
 
\begin{deluxetable}{lrr}
\tablecaption{\label{tb:highmass}The best examples of fitted DAs with
likely high \logg values above 8.5.} 
\tablecomments{The clustering of \logg around 9.0
(with small errors) is an artifact of our limited model grid and our
fitting technique at the limits of our grid. The true \logg values are
likely high, but may not be exactly as determined here.}
\tablewidth{0pt}
\tablehead{ \colhead{Name} & \colhead{\logg} & \colhead{\Teff}}
\startdata
SDSS~J144707.42+585506.7  & $8.94 \pm 0.06$ &  $14802 \pm 439$ \\
SDSS~J024700.48--070547.1 & $8.97 \pm 0.04$ &  $19866 \pm 791$ \\
SDSS~J113509.97+642949.0  & $8.98 \pm 0.03$ &  $9031 \pm 34$ \\
SDSS~J002049.39+004435.1  & $9.00 \pm 0.00$ &  $9182 \pm 17$ \\
SDSS~J011055.07+143922.3  & $9.00 \pm 0.00$ &  $9406 \pm 18$ \\
SDSS~J020626.89--005710.0 & $9.00 \pm 0.01$ &  $7841 \pm 37$ \\
SDSS~J075916.54+433519.1  & $9.00 \pm 0.00$ &  $22222 \pm 392$ \\
SDSS~J155238.21+003910.4  & $9.00 \pm 0.00$ &  $16981 \pm 441$ \\
\enddata
\end{deluxetable}

\clearpage

\begin{deluxetable}{llllll}
\tablecaption{\label{tb:known}Previously known white dwarf stars recovered in
this work.}


\tabletypesize{\small}
\tablewidth{0pt}
\tablehead {
\colhead{SDSS Name} & \colhead{g$_{psf}$} & \colhead{SDSS Type} & 
\colhead{Other Name} & \colhead{Other Type} & \colhead{Note} }
\startdata
SDSS~J001339.11+001924.9 &  15.37 & DA5.3:  & WD0011+000  & DA6    &  1\\
SDSS~J002602.29--103752.0 &  16.22 & DA5.0:  & WD0023--109  & DA7    &  1\\
SDSS~J003230.11+001138.4 &  18.64 & DB3.7:  & WD0029--000  & DB\\
SDSS~J003508.26+135045.3 &  16.37 & DA2.3:  & HS0032+1334 & DA     &  2\\
SDSS~J003426.93+151801.8 &  16.99 & DA6.6:  & WD0031+150  & DA7\\
SDSS~J004022.88--002130.1 &  14.83 & DA3.3:  & WD0037--006  & DA4    &  1\\
SDSS~J010207.17--003259.5 &  18.21 & DA4.6   & WD0059+002  & DA\\
SDSS~J011009.10+132616.1 &  16.54 & DAMe    & HS0107+1310 & DA     &  2\\
SDSS~J011055.07+143922.3 &  16.91 & DA5.3   & WD0108+143  & DA     &  1\\
SDSS~J021028.69+124319.0 &  16.86 & DA3.0:  & HS0207+1229 & DA     &  2\\
SDSS~J024602.67+002539.3 &  17.21 & DA3.5:  & WD0243+002  & DA     &  1\\
SDSS~J024821.95+005109.1 &  17.99 & DA3.3   & WD0245+006  & DA\\
SDSS~J025200.98+004544.2 &  18.41 & DA4.9   & WD0249+005  & DA     &  1\\
SDSS~J025624.74+003558.0 &  18.07 & DA1.3   & WD0253+003  & DA\\
SDSS~J025709.00+004628.1 &  17.38 & DA4.1   & WD0254+005  & DA\\
SDSS~J025746.41+010106.0 &  17.66 & DA3.0   & WD0255+008  & DA\\
SDSS~J025801.20--005400.1 &  18.03 & DA5.3   & WD0255--010  & DA\\
SDSS~J025817.87+010946.0 &  18.20 & DAM     & WD0255+009.2 & DA\\
SDSS~J030407.40--002541.7 &  17.75 & DAH     & WD0301--006  & DAH3.4 &  3\\
SDSS~J031305.82--070749.5 &  16.47 & DA2.8   & WD0310--073  & DA\\
SDSS~J032302.85+000559.7 &  17.44 & DA3.8   & WD0320--000  & DA\\
SDSS~J033133.89+010327.9 &  16.43 & DA1.4:  & WD0328+008  & DA\\
SDSS~J033145.69+004517.0 &  17.21 & DAH     & WD0329+005  & DAH\\
SDSS~J033200.49--005752.5 &  17.07 & DA2.9   & WD0329--011  & DA\\
SDSS~J033320.37+000720.7 &  16.53 & DBH     & WD0330--000  & DB:HP\\
SDSS~J034511.11+003444.3 &  18.63 & DH      & WD0342+004  & DAH6.3 &  3\\
SDSS~J075723.93+400714.8 &  17.55 & DA2.6   & WD0754+402  & DA\\
SDSS~J075959.56+433521.3 &  16.19 & DAH     & WD0756+437  & DAH    &  1\\
SDSS~J080459.02+415744.9 &  17.45 & DA3.6   & WD0801+421  & DA\\
SDSS~J084951.11+553514.7 &  16.20 & DA1.8:  & WD0846+557  & DA2\\
SDSS~J093958.66+011638.2 &  16.45 & DA2.6:  & HS0937+0130 & DA     &  2\\
SDSS~J094640.35+011319.9 &  17.18 & DA2.5:  & HS0944+0127 & DA     &  2\\
SDSS~J095102.23+010432.6 &  15.59 & DB2.9:  & WD0948+013  & DB2\\
SDSS~J095220.45+005913.2 &  18.96 & DA4.8   & WD0949+012  & DA5.0\\
SDSS~J095245.59+020938.9 &  16.35 & DA1.2:  & WD0950+023  & DA1    &  1\\
SDSS~J095810.68--010417.8 &  16.51 & DA2.1:  & WD0955--008  & DA2\\
SDSS~J100316.35--002337.0 &  15.96 & DA2.5:  & WD1000--001  & DA2.5\\
SDSS~J101219.90+004019.7 &  17.72 & DQ      & WD1009+009  & DC     &  1\\
SDSS~J101232.49+015444.6 &  18.04 & DA2.2:  & WD1009+021  & DA2.0\\
SDSS~J101548.01+030648.4 &  15.66 & DA4.3:  & HS1013+0321 & DA     &  2\\
SDSS~J101607.40+002038.2 &  18.71 & DA2.5   & WD1013+005  & DA2.5\\
SDSS~J101805.04+011123.5 &  16.29 & DAH     & WD1015+014  & DAP3.5\\
SDSS~J102549.72+003906.2 &  16.07 & DA1.4:  & WD1023+009  & DA1.5  &  1\\
SDSS~J102732.54--005440.1 &  18.76 & DA2.0   & WD1024--006  & DA2.5\\
SDSS~J103004.51--010919.1 &  18.71 & DA1.9:  & WD1027--008  & DA2.0\\
SDSS~J103448.94+005201.3 &  19.08 & DA5.2:  & WD1032+011  & DA3.0\\
SDSS~J103635.66--000036.4 &  18.92 & DA3.6:  & WD1034+002  & DA3.5\\
SDSS~J104946.47+003635.1 &  17.25 & DA2.2   & HS1047+0052 & DA     &  2\\
SDSS~J110515.32+001626.1 &  15.20 & DA3.9:  & HS1102+0032 & DA     &  2\\
SDSS~J110636.72--001122.4 &  18.32 & DA3.3   & WD1104+000  & DA3.5\\
SDSS~J111028.70--003343.5 &  18.61 & DA5.2:  & WD1107--002  & DA5.0\\
SDSS~J113901.22+000321.8 &  18.87 & DA3.7   & WD1136+003  & DA3.0\\
SDSS~J114312.57+000926.5 &  18.15 & DAM     & WD1140+004  & DA4.0+M\\
SDSS~J114425.06+013949.4 &  18.19 & DA4.0   & WD1141+019  & DA\\
SDSS~J114635.23+001233.4 &  14.88 & PG1159  & WD1144+004  & DQZO1\\
SDSS~J114913.53--014728.6 &  17.98 & DAM     & WD1146--015  & DA     &  1\\
SDSS~J115418.14+011711.4 &  17.75 & DAH     & HS1151+0133 & DA     &  2\\
SDSS~J121635.37--002656.2 &  19.60 & DAH     & WD1214--001  & DAH    &  3\\
SDSS~J122209.44+001534.0 &  20.27 & DAH     & SDSS J1222  & DAH    &  3\\
SDSS~J123706.24--001603.9 &  19.05 & DA3.1   & WD1234+000  & DA3.5\\
SDSS~J123819.77+005248.2 &  18.91 & DA5.6   & WD1235+011  & DA\\
SDSS~J123836.35--004042.3 &  17.41 & DAM     & WD1236--004  & DA\\
SDSS~J123836.74--013936.2 &  18.84 & DA3.2   & WD1236--013  & DA     &  1\\
SDSS~J123910.18--010005.4 &  19.08 & DA1.9   & WD1236--007  & DA     &  1\\
SDSS~J123922.34+005548.8 &  19.27 & DAM     & WD1236+012  & DA3.0+M\\
SDSS~J124438.81--022107.5 &  18.39 & DA4.6   & WD1242--020  & DA3.5\\
SDSS~J124709.83+005533.0 &  19.37 & DA2.7   & WD1244+011  & DA4.0\\
SDSS~J124920.09+001911.6 &  19.73 & DA2.2:  & WD1246+005  & DA2.5\\
SDSS~J125139.78--010254.1 &  18.38 & DA4.9   & WD1249--007  & DA3.0\\
SDSS~J125730.31--001150.9 &  18.79 & DA1.7   & WD1254+000  & DA1.5\\
SDSS~J130110.51+010739.9 &  16.30 & DA4.5   & WD1258+013  & DA\\
SDSS~J130815.22--015904.5 &  16.80 & DA0.9   & WD1305--017  & DAO1   &  1\\
SDSS~J131717.02--021945.6 &  18.32 & DB2.8   & WD1314--020  & DB\\
SDSS~J131724.75+000237.4 &  15.77 & DO      & WD1314+003  & DO\\
SDSS~J132232.12+641545.8 &  16.25 & DA1.8:  & WD1320+645  & DA2    &  1\\
SDSS~J132439.71--031923.5 &  18.15 & DA3.5   & WD1322--030  & DA     &  1\\
SDSS~J133137.06+010632.1 &  17.43 & DA1.4   & HS1329+0121 & DA     &  2\\
SDSS~J133739.40+000142.9 &  19.57 & DC      & WD1335+002  & DC     &  4\\
SDSS~J133838.48--000712.4 &  18.70 & DA4.9   & WD1336+001  & DA3.0\\
SDSS~J134430.11+032423.2 &  16.61 & DA3.7:  & HS1341+0339 & DA     &  2\\
SDSS~J135211.00+652457.1 &  15.44 & DA4.2:  & WD1350+656  & DAV4.2\\
SDSS~J135459.89+010819.3 &  16.36 & DA4.3:  & HS1352+0123 & DA     &  2\\
SDSS~J135532.42+001124.0 &  15.71 & DB3.2:  & WD1352+004  & DB4\\
SDSS~J141011.44+045255.8 &  17.40 & DA3.4   & HS1407+0507 & DA     &  2\\
SDSS~J141457.89+012207.4 &  17.83 & DA5.5:  & HS1412+0136 & DA     &  2\\
SDSS~J143947.62--010606.9 &  16.52 & DAMe    & WD1437--008  & DC\\
SDSS~J144433.80--005958.9 &  16.22 & DA4.0   & WD1441--007  & DA3\\
SDSS~J144518.03+585032.2 &  17.70 & DBZ     & WD1443+590  & DB\\
SDSS~J144828.21--010525.5 &  18.87 & DA3.8   & WD1445--008  & DA3.5\\
SDSS~J145535.49+010246.5 &  18.95 & DA5.4   & WD1453+012  & ...\\
SDSS~J145600.81+574150.8 &  16.19 & DA1.6:  & WD1454+578  & DA\\
SDSS~J145644.91+011017.6 &  19.05 & DB3.1   & WD1454+013  & ...\\
SDSS~J145947.04--003954.6 &  18.40 & DB3.0:  & WD1457--004  & ...\\
SDSS~J150003.86+002420.0 &  18.80 & DB3.4   & WD1457+006  & ...\\
SDSS~J150231.66+011045.9 &  18.47 & DAM     & WD1459+013  & ...\\
SDSS~J150547.49+024840.6 &  16.34 & DA2.8   & HS1503+0300 & DA     &  2\\
SDSS~J151151.36+562450.5 &  16.31 & DA5.5   & WD1510+566  & DA6    &  1\\
SDSS~J151421.26+004752.8 &  15.68 & DA1.8   & WD1511+009  & DA2\\
SDSS~J152839.42+011300.1 &  16.45 & DA0.9   & WD1526+013  & DA1\\
SDSS~J154338.69+001202.1 &  16.76 & sdB     & WD1541+003  & DAwk    & 1\\
SDSS~J165401.26+625355.0 &  18.40 & DC      & WD1653+630  & DC9\\
SDSS~J165935.58+620933.9 &  16.25 & DA4.1   & WD1659+622  & DA\\
SDSS~J172045.37+561214.9 &  20.10 & DAH     & WD1719+562  & DAH     &  3\\
SDSS~J172329.14+540755.8 &  18.78 & DAH     & WD1722+541  & DAH3.1  &  3\\
SDSS~J172643.38+583732.2 &  15.32 & DA0.8:  & WD1725+586  & DA      &  1\\
SDSS~J172856.22+555822.8 &  15.98 & DQABCI  & WD1727+560  & DQAB?4\\
SDSS~J232248.22+003900.9 &  19.14 & DAH     & WD2320+003  & DAH1.3  &  3\\
SDSS~J232337.55--004628.2 &  17.98 & DBH     & WD2321--010  & DAH?2.5 &  3\\
SDSS~J235410.39--010728.5 &  18.19 & DB3.5:  & WD2351--014  & DB      &  1\\
\enddata

\tablenotetext{1}{SDSS position different from previous by more than 10 arcsec.}
\tablenotetext{2}{Unpublished white dwarf, in Hamburg Quasar Survey (Homeier 2002, priv. comm.).}
\tablenotetext{3}{White dwarf discovered in the SDSS EDR (Gansicke et al. 2002).}
\tablenotetext{4}{White dwarf discovered in the SDSS (Harris et al. 2001).}

\end{deluxetable}

\begin{deluxetable}{llrrrrrl}
\tablecaption{\label{tb:notindr1}Previously known white dwarf stars not 
spectroscopically recovered in this work.}
\tablecomments{Table~\ref{tb:notindr1} is published in its entirety in the
electronic edition of the {\it Astrophysical Journal} and at the URL provided
in the text. A portion is shown here
for guidance regarding its form and content.  The SDSS 
{\it g,r,i,z} magnitudes have been removed from this sample table which
displays the data from the first and last three entries of the full table.}
\tablewidth{0pt}
\tablehead{
\colhead{Name} & \colhead{Type} & \colhead{Mag} &\colhead{RA} &
\colhead{Dec} &
\colhead{u$_{psf}$} & \colhead{$\delta_u$} &
\colhead{Notes}\\
\colhead{} & \colhead{} & \colhead{} & \colhead{(J2000)} & \colhead{(J2000)}
}
\startdata
WD0041--102 & DBAP3 & 14.47~v & 00~43~45.98 & --10~00~25.1 & 14.25 & 0.01                 &\\
WD0042+140 & DC: & 18.9~p & 00~45~25.79 & +14~21~29.4 & 22.00 & 0.25                 &\\
WD0106--109.1 & DA & 16.5~p & 01~09~03.43 & --10~42~14.2 & 17.15 & 0.02                 &\\
WD2318+007.1 & DC: & 18.8~p & 23~21~15.32 & +01~02~11.3 & 20.53 & 0.10                 &\\
WD2318+007.2 & DC: & 19.7~p & 23~21~15.68 & +01~02~23.9 & 21.70 & 0.26                 &\\
WD2333--002 & DA2? & 15.49~p & 23~35~41.47 & +00~02~19.5 & 15.30 & 0.02                 &\\
\enddata

\tablenotetext{*}{Star image contains saturated pixels in this filter.}
\tablenotetext{1}{No white dwarf found near this position.}
\tablenotetext{2}{SDSS spectrum 0410--51816--565 shows WD0255+009.1 is a QSO.}
\tablenotetext{3}{No SDSS imaging data, too near bright star.}
\tablenotetext{4}{Probably same star as WD0330--009.}
\tablenotetext{5}{WD0820+021 probably not detected, too faint.}
\tablenotetext{6}{Spectrum not in DR1, but is given in Initial Survey paper
(Harris et al. 2003).}
\tablenotetext{7}{Falls in small gap in SDSS imaging data.}
\tablenotetext{8}{Unresolved with WD1330+015.1.}
\tablenotetext{9}{The colors of WD1401+005 are red and indicate it is 
                  not a white dwarf.}
\tablenotetext{10}{Probably same star as WD1422+028.}
\tablenotetext{11}{Colors are very red and indicate WD1449+003 is not a white
                   dwarf.}
\tablenotetext{12}{Colors indicate WD1451--004, WD1455+019, and WD1500+006 are
                   horizontal-branch stars, not white dwarf stars.}
\tablenotetext{13}{Probably same star as WD1544+009.}

\end{deluxetable}

\clearpage

\end{document}